\documentclass[]{aa}  
\pdfoutput=1
\usepackage{graphicx}
\usepackage{caption}
\usepackage{subcaption}
\usepackage{epsf}
\usepackage{longtable,lscape}
\usepackage{dcolumn}
\usepackage{booktabs} 
\usepackage{url}
\usepackage{float}
\usepackage{microtype}
\usepackage[varg]{txfonts}
\usepackage{natbib,twoopt} 
\usepackage{longtable}
\usepackage{lipsum}
\def\apjl{ApJ}%
\def\apjs{ApJS}%
\def\ao{Appl.~Opt.}%
\def\aap{A\&A}%
\bibpunct{(}{)}{;}{a}{}{,} 

\let\oldhat\hat

\renewcommand{\hat}[1]{\oldhat{\mathbf{#1}}}
\newcommand{\kms}{km\,${\rm s}^{-1}\,$}

\titlerunning{A first SB2 orbital and spectroscopic analysis for the Wolf-Rayet binary R145}

\begin{document}

   \title{The Tarantula Massive Binary Monitoring project: II. A first SB2 orbital and spectroscopic analysis for the Wolf-Rayet binary R145}


   \author{T.\ Shenar\inst{1} 
          \and N.\ D.\ Richardson\inst{2}      
          \and D.\ P.\ Sablowski\inst{3}
          \and R.\ Hainich\inst{1}
          \and H.\ Sana\inst{4}    
          \and A.\ F.\ J.\ Moffat\inst{5}
          \and H.\ Todt\inst{1}
          \and W.-R.\ Hamann\inst{1}         
          \and L.\ M.\ Oskinova\inst{1}
          \and A.\ Sander\inst{1}
          \and F.\ Tramper\inst{6}
          \and N.\ Langer\inst{7}
          \and A.\ Z.\ Bonanos\inst{8}
          \and S.\ E.\ de Mink\inst{9}
          \and G.\ Gr\"afener\inst{10}
          \and P.\ A.\ Crowther\inst{11}
          \and J.\ S.\ Vink\inst{10}
          \and L.\ A.\ Almeida\inst{12,13} 
          \and A.\ de Koter\inst{4,9}
          \and R.\ Barb\'a\inst{14}
          \and A.\ Herrero\inst{15}
          \and K.\ Ulaczyk\inst{16}
          }
          
   \institute{\inst{1}{Institut f\"ur Physik und Astronomie, Universit\"at Potsdam,
                Karl-Liebknecht-Str. 24/25, D-14476 Potsdam, Germany}\\
              \email{shtomer@astro.physik.uni-potsdam.de}  \\                
              \inst{2}{Ritter Observatory, Department of Physics and Astronomy, The University of Toledo, Toledo, OH 43606-3390, USA}\\
              \inst{3}{Leibniz-Institut f\"ur Astrophysik Potsdam, An der Sternwarte 16, 14482 Potsdam, Germany}\\
              \inst{4}{Institute of Astrophysics, KU Leuven, Celestijnenlaan 200 D, 3001, Leuven, Belgium} \\              
              \inst{5}{D\'epartement de physique and Centre de Recherche en Astrophysique 
                du Qu\'ebec (CRAQ), Universit\'e de Montr\'eal, C.P. 6128, Succ.~Centre-Ville, Montr\'eal, Qu\'ebec, H3C 3J7, Canada}\\   
              \inst{6}{European Space Astronomy Centre (ESA/ESAC), PO Box 78, 28691 Villanueva de la Ca\~nada, Madrid, Spain} \\    
              \inst{7}{Argelander-Institut f\"ur Astronomie der Universit\"at Bonn, Auf dem H\"ugel 71, 53121 Bonn, Germany} \\                  
              \inst{8}{IAASARS, National Observatory of Athens, GR-15236 Penteli, Greece} \\                  
              \inst{9}{Anton Pannenkoek Astronomical Institute, University of Amsterdam, 1090 GE Amsterdam, Netherlands} \\   
              \inst{10}{Armagh Observatory, College Hill, Armagh, BT61 9DG, Northern Ireland, UK} \\                
              \inst{11}{Department of Physics and Astronomy, University of Sheffield, Sheffield S3 7RH} \\                      
              \inst{12}{Instituto de Astronomia, Geof\'isica e Ci\^{e}ncias, Rua do Mat\~{a}o 1226, Cidade Universit\'{a}ria S\~{a}o Paulo, SP, Brasil} \\      
              \inst{13}{Dep.\ of Physics \& Astronomy, Johns Hopkins University, Bloomberg Center for Physics and Astronomy, 3400N Charles St, USA} \\
              \inst{14}{Departamento de Física y Astronom\'ia, Universidad de la Serena, Av. Juan Cisternas 1200 Norte, La Serena}\\ 
              \inst{15}{Instituto de Astrof\'isica, Universidad de La Laguna, Avda.\ Astrof\'isico Francisco S\'anchez s/n, E-38071 La Laguna, Tenerife, Spain} \\    
              \inst{16}{Warsaw University Observatory, Al. Ujazdowskie 4, 00-478 Warszawa, Poland} \\                     
              }
 \date{Received August 31, 206 / Accepted October 20, 2016}


\abstract
{
We present the first SB2 orbital solution and disentanglement of the massive Wolf-Rayet binary 
R145 ($P = 159\,$d) located in the Large Magellanic Cloud.  
The primary was claimed to have a stellar mass 
greater than $300\,M_\odot$, making it a candidate for the most massive star known.
While the primary is a known late type, H-rich Wolf-Rayet star (WN6h), the secondary could not be so far unambiguously detected. 
Using moderate resolution spectra, we are able to derive accurate radial velocities 
for both components. By performing simultaneous orbital and polarimetric analyses, we derive 
the complete set of orbital parameters, including the inclination.
The spectra are disentangled and spectroscopically analyzed, and an analysis of the wind-wind collision 
zone is conducted. \\
The disentangled spectra and our models are consistent with a WN6h type for the primary, and suggest that 
the secondary is an O3.5~If*/WN7 type star.
We derive a high eccentricity of $e = 0.78$ and 
minimum masses of $M_1 \sin^3 i \approx M_2 \sin^3 i = 13\pm2\,M_\odot$, with $q = M_2 / M_1 = 1.01\pm0.07$. 
An analysis of emission excess stemming from a wind-wind collision
yields a similar inclination to that obtained from polarimetry ($i = 39\pm6^\circ$).
Our analysis thus implies $M_1 = 53^{+40}_{-20}$ and $M_2 = 54^{+40}_{-20}\,M_\odot$, excluding $M_1 > 300\,M_\odot$. A detailed comparison with evolution tracks calculated for 
single and binary stars, as well as the high eccentricity, suggest that the components of the system underwent quasi-homogeneous evolution and avoided mass-transfer. 
This scenario would suggest current masses of $\approx 80\,M_\odot$ and initial masses of $M_\text{i, 1} \approx 105$ and $M_\text{i,2} \approx 90\,M_\odot$, 
consistent with the upper limits of 
our derived orbital masses, and would imply an age of $\approx 2.2\,$Myr.  
}


\maketitle

\section{Introduction}
\label{sec:introduction}


There are ever growing efforts to discover the most massive stars in the Universe
\citep[e.g.,][]{Massey1998, SchnurrNGC2008, Bonanos2009, Bestenlehner2011, Tramper2016}.
Because of their extreme influence on their environment, understanding the formation, 
evolution, and death of massive stars is imperative for a multitude of astrophysical fields. 
Establishing the upper mass limit for stars is one of the holy grails of stellar physics, laying sharp constraints 
on the initial mass function \citep{Salpeter1955, Kroupa2001} and massive
star formation \citep{Bonnell1997, Oskinova2013b}. Current estimates for an upper mass limit range from $\approx 120\,M_\odot$ 
\citep[e.g.,][]{Oey2005} to $\gtrsim 300\,M_\odot$ \citep[e.g.,][]{Crowther2010, Schneider2014b, Vinkbook}. 

However, the only reliable method to weigh stars is by analyzing the orbits of stars in binaries \citep{Andersen1991, Torres2010}. 
This is especially crucial for massive Wolf-Rayet (WR) stars, whose powerful winds make it virtually impossible to
estimate their surface gravities. 
Fortunately, massive stars ($M \gtrsim 8\,M_\odot$) 
tend to  exist in binary or multiple systems \citep{Mason2009, Maiz2010, Oudmaijer2010, Sana2012, Sana2014, Sota2014, Aldoretta2015}. 

%
Primarily due to mass-transfer, the evolutionary path of a star in a binary can greatly deviate from that of an identical star 
in isolation \citep{Paczynski1973, Langer2012, DeMink2014}. 
The impact and high frequency of binarity make binaries both indispensable laboratories for the study of massive stars, as well as 
important components of stellar evolution. Hence, it is imperative to discover and study 
massive binary systems in the Galaxy and the Local Group. 

\begin{figure}[!htb]
\centering
  \includegraphics[width=\columnwidth]{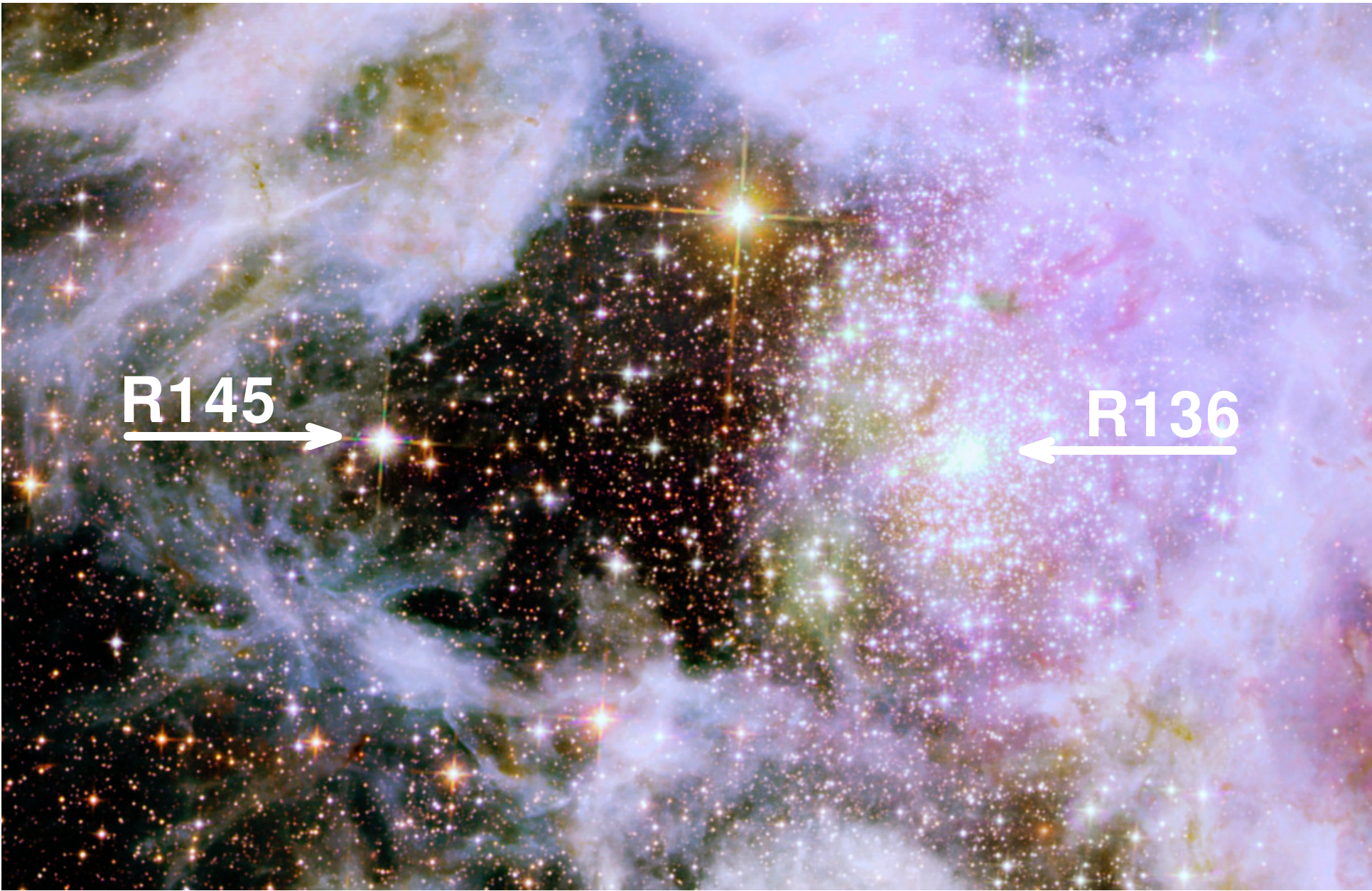}
  \caption{The area of sky around R145 (Credit: NASA, ESA, E. Sabbi, STScI). The image was obtained
using the Hubble Space Telescope's (HST) WFC3 and ACS cameras in filters which roughly overlap with the I, J, and H bands. The image size is $\approx 2.5'\times 1.7'$. North is up and east to the
left. The arrows indicate the positions of R145 and the star cluster R\,136. The distance between R145 and R\,136 is $\approx
1.3'$ or in projection $\approx 19$\,pc.
}
\label{fig:R136}
\end{figure}

\object{R145} (\object{BAT99 119}, \object{HDE 269928}, \object{Brey 90}, \object{VFTS 695}) is a known massive binary situated in the famous Tarantula nebula 
in the Large Magellanic Cloud (LMC), about $19\,$pc away from the massive cluster \object{R\,136} in projection (see Fig.\,\ref{fig:R136}).
The system's composite spectrum was classified WN6h in the original BAT99 catalog \citep{Breysacher1999},  
which, according to this and past studies \citep[e.g.,][S2009 hereafter]{Schnurr2009}, corresponds to the primary.
The primary thus belongs to the class of late WR stars which have not yet exhausted their hydrogen 
content, and is likely still core H-burning.

The system was speculated to host some of the most massive and luminous 
stars in the Local Group. Erroneously assuming a circular orbit, 
\citet{Moffat1989} detected a periodic Doppler shift with a period of $P = 25.17\,$d, concluding R145 to be an SB1 binary.
A significantly different period of $158.8\pm0.1$\,d was later reported by 
S009, who combined data from \citet{Moffat1989} with their own and found a highly eccentric system. 
S009 could not derive a radial velocity (RV) curve for the secondary. However, they attempted to estimate the secondary's RV amplitude 
by looking for ``resonance'' velocity amplitudes which would strengthen the secondary's features in a spectrum formed by co-adding the 
spectra in the secondary's frame of reference, 
under the assumption that the secondary is mainly an 
absorption-line star moving in anti-phase to the primary star. Combined with an orbital inclination of $i = 38^\circ\pm9$ they derived, 
their results tentatively implied that the system comprises two incredibly massive stars of $\approx 300$ and $\approx 125\,M_\odot$, making it potentially the most 
massive binary system known. For comparison, binary components of similar spectral 
type typically have masses ranging from $\approx 50$ to $\approx 100\,M_\odot$ 
(e.g.,\ \object{WR\,22}, \citealt{Rauw1996}; \object{WR\,20a}, 
\citealt{Bonanos2004}, \citealt{Rauw2004}; \object{WR\,21a}, \citealt{Niemela2008}, \citealt{Tramper2016}; \object{HD 5980}, \citealt{Koenigsberger2014}; \object{NGC 3603-A1}, \citealt{SchnurrNGC2008}); 
masses in excess of $300\,M_\odot$ 
were so far only reported for putatively single stars \citep[e.g.,][]{Crowther2010} based on comparison with evolutionary models.

With such high masses, signatures for wind-wind collisions (WWC) are to be expected \citep{Moffat1998}. WWC excess emission 
can be seen photometrically as well as spectroscopically, and can thus also introduce a bias when deriving RVs.
WWC signatures do not only reveal
information on the dynamics and kinematics of the winds, but can also constrain the orbital inclination, which is crucial for an accurate 
determination of the stellar masses. 
Polarimetry offers a further independent tool to constrain orbital inclinations of binary systems \citep{Brown1978}. 
Both approaches are used in this study to constrain 
the orbital inclination $i$. For high inclination angles, photometric variability due to photospheric/wind eclipses 
can also be used to constrain $i$ \citep[e.g.,][]{Lamontagne1996}. 
At the low inclination angle of R145 (see Sects.\,\ref{sec:pol} and \ref{sec:wwc}, as well as S2009), however, eclipses are not expected to 
yield significant constraints.

Using $110$ high-quality Fibre Large Array Multi Element Spectrograph
(FLAMES) spectra (Sect.\,\ref{sec:obsdata}), we are able
to derive for the first time  a double-lined spectroscopic orbit for R145. We  identify lines which enable us to construct a 
reliable SB2 RV curve (Sect.\,\ref{sec:RVmes}). 
The majority of the spectra were taken as part of the VLT FLAMES-Tarantula survey \citep[VFTS,][]{Evans2011}
and follow up observations (PI: H.\ Sana). 
The study is conducted in the framework of the Tarantula Massive Binary Monitoring (TMBM) project \citep[][submitted to A\&A, paper I hereafter]{Almeida2016}.

The RVs of both components are fitted 
simultaneously with polarimetric data to obtain accurate orbital parameters (Sect.\,\ref{sec:pol}).
In Sect.\,\ref{sec:disentangle}, we disentangle the spectrum to its constituent spectra.
Using the disentangled spectra, an XSHOOTER spectrum, and additional observational material, 
we perform a multiwavelength spectroscopic analysis of the system using the Potsdam Wolf-Rayet ({\sc PoWR}) 
model atmosphere code to derive the fundamental stellar parameters
and abundances of both stars (Sect.\,\ref{sec:specan}). An analysis of WWC signatures is presented in Sect.\,\ref{sec:wwc}.  
A discussion of the evolutionary status of the system in light of our results  is presented in Sect.\,\ref{sec:disc}. We conclude 
with a summary in Sect.\,\ref{sec:summary}.

\section{Observational data}
\label{sec:obsdata}

The FLAMES spectra
(072.C-0348, Rubio; 182.D-0222, Evans; 090.D-0323, Sana; 092.D-0136, Sana)
were secured between 2004 and 2014 with the FLAMES instrument mounted on the Very Large Telescope (VLT), Chile, partly in the 
course of two programs: the VLT FLAMES Tarantula Survey \citep{Evans2011} and the TMBM project. 
They cover the spectral range $3960 - 4560\,\AA$, have a typical signal-to-noise (S/N) ratio of 100, and a resolving power of $R \approx 8000$
(see paper I for more information). 
The spectra are rectified 
using an automated routine which fits a piecewise first-order polynomial to the apparent continuum, and are cleaned from cosmic events using a self-written 
Python routine.

\begin{figure}[!htb]
\centering
  \includegraphics[width=\columnwidth]{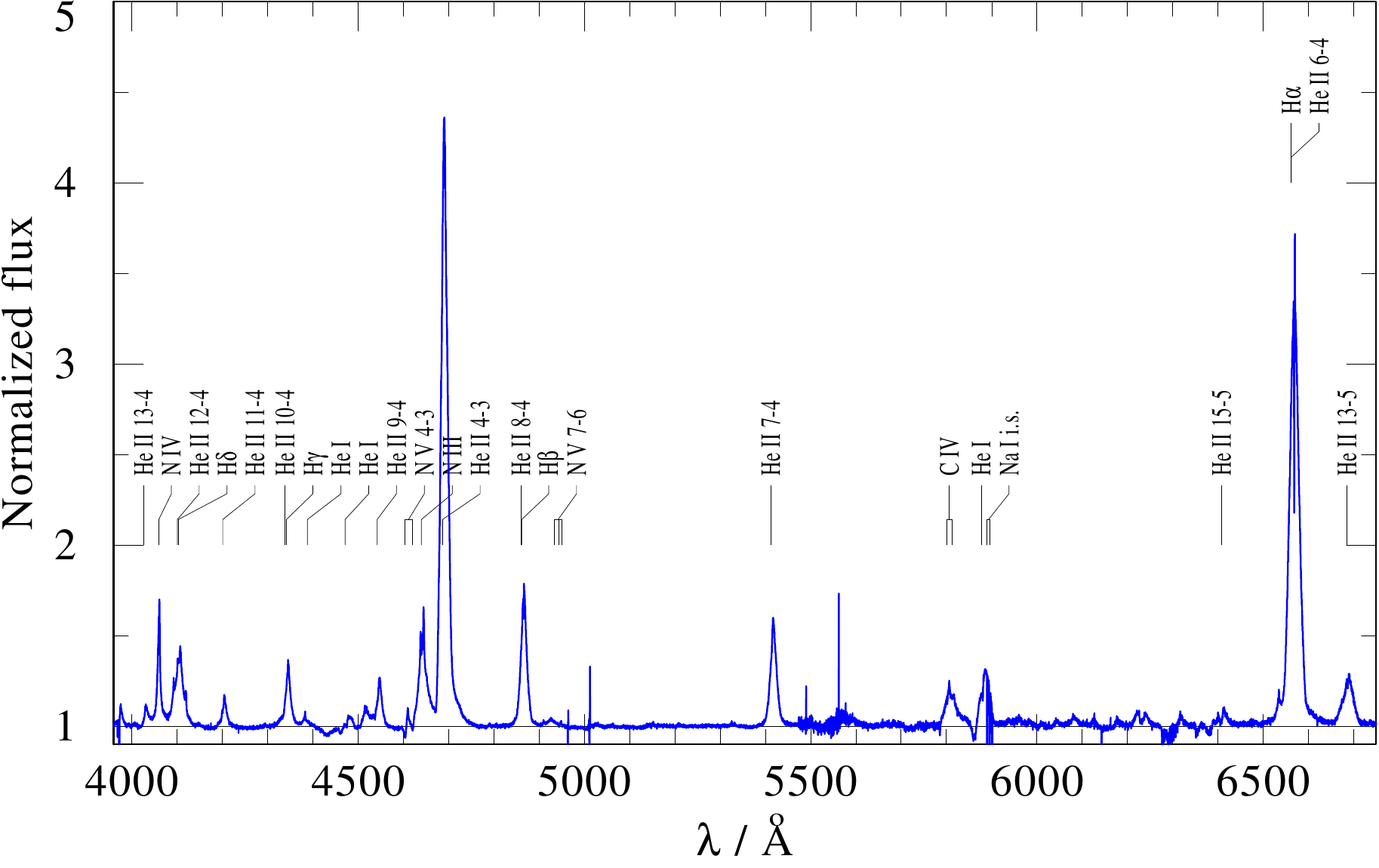}
  \caption{Segment of the XSHOOTER spectrum of R145.} 
\label{fig:specoverview-xshooter}
\end{figure}

For the spectral analysis, we use an XSHOOTER \citep{Vernet2011} spectrum 
(085.D-0704, PI: Sana) taken on 22 April 2010 ($\phi \approx 0.5$, i.e.\ apastron, 
with the phase calculated using the ephemeris given in Table\,\ref{tab:orbpar} in Sect.\,\ref{sec:pol}).
The spectrum covers the range 
$3000 - 25000\,\AA$. 
It has $\text{S/N} \approx 100$ and a resolving power
of $R \approx 7500$ in the spectral range $5500 - 10000\,\AA$ and $R \approx 5000$ in the 
ranges $3000 - 5500\,\AA$ and $10000 - 25000\,\AA$. It is rectified by fitting a first-order polynomial to the apparent continuum. 
A segment of the spectrum is shown in Fig.\,\ref{fig:specoverview-xshooter}.

We make use of two high-resolution (HIRES), flux-calibrated International Ultraviolet Explorer (IUE) spectra available in the IUE archive. 
The two spectra (swp47847, PM048, PI: Bomans; swp47836, PM033, PI: de Boer)  were obtained on 08 and 10 June 1993 
at roughly $\phi = 0.7$ and are thus 
co-added to enhance the S/N. The co-added spectrum has a resolving power of $R \approx 10000$ and $\text{S/N} \approx 10$. The spectrum is rectified 
using the composite {\sc PoWR} model continuum (see Sect.\,\ref{sec:specan}).

Linear polarimetry was obtained between 1988 and 1990 at the 2.15-m ``Jorge Sahade'' telescope of the Complejo Astron\'omico El Leoncito (CASLEO) 
near San Juan, Argentina,
with the Vatican Observatory Polarimeter \citep[VATPOL,][]{Magalhaes1984} and at the European 
Southern Observatory (ESO)/Max-Planck-Gesellschaft (MPG)-2.2m 
telescope at La Silla, Chile, with Polarimeter with Instrumentation and Sky COmpensation \citep[PISCO,][]{Stahl1986}. The data were obtained and originally used by S2009, 
where more details can be found.

\section{SB2 orbit construction}
\label{sec:RVmes}

As discussed in the introduction, R145 is a known SB1 binary. However, no previous studies could unambiguously 
isolate the secondary in the spectrum. Thanks to the moderate resolution, high S/N of the FLAMES spectra, we are now able 
to detect spectral features which belong solely to the secondary
and thus construct an SB2 orbit for the system.

\begin{figure}[!htb]
\centering
  \includegraphics[width=\columnwidth]{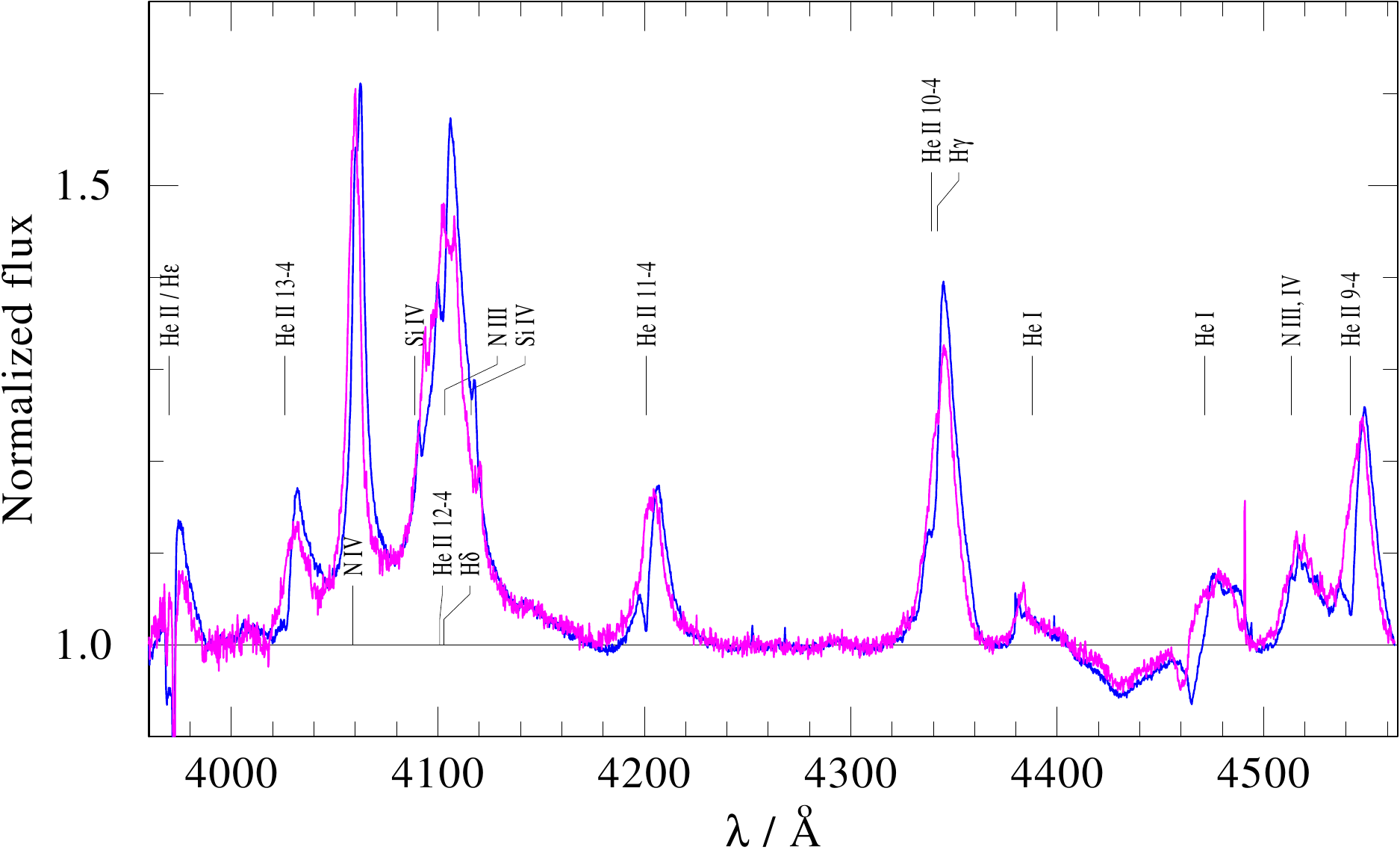}
  \caption{Two FLAMES spectra taken slightly before ($\phi = -0.03$) and after ($\phi = 0.07$) periastron passage 
  (extreme velocity amplitudes).}
\label{fig:specoverview-flames}
\end{figure}

\begin{figure}[!htb]
\centering
  \includegraphics[width=\columnwidth]{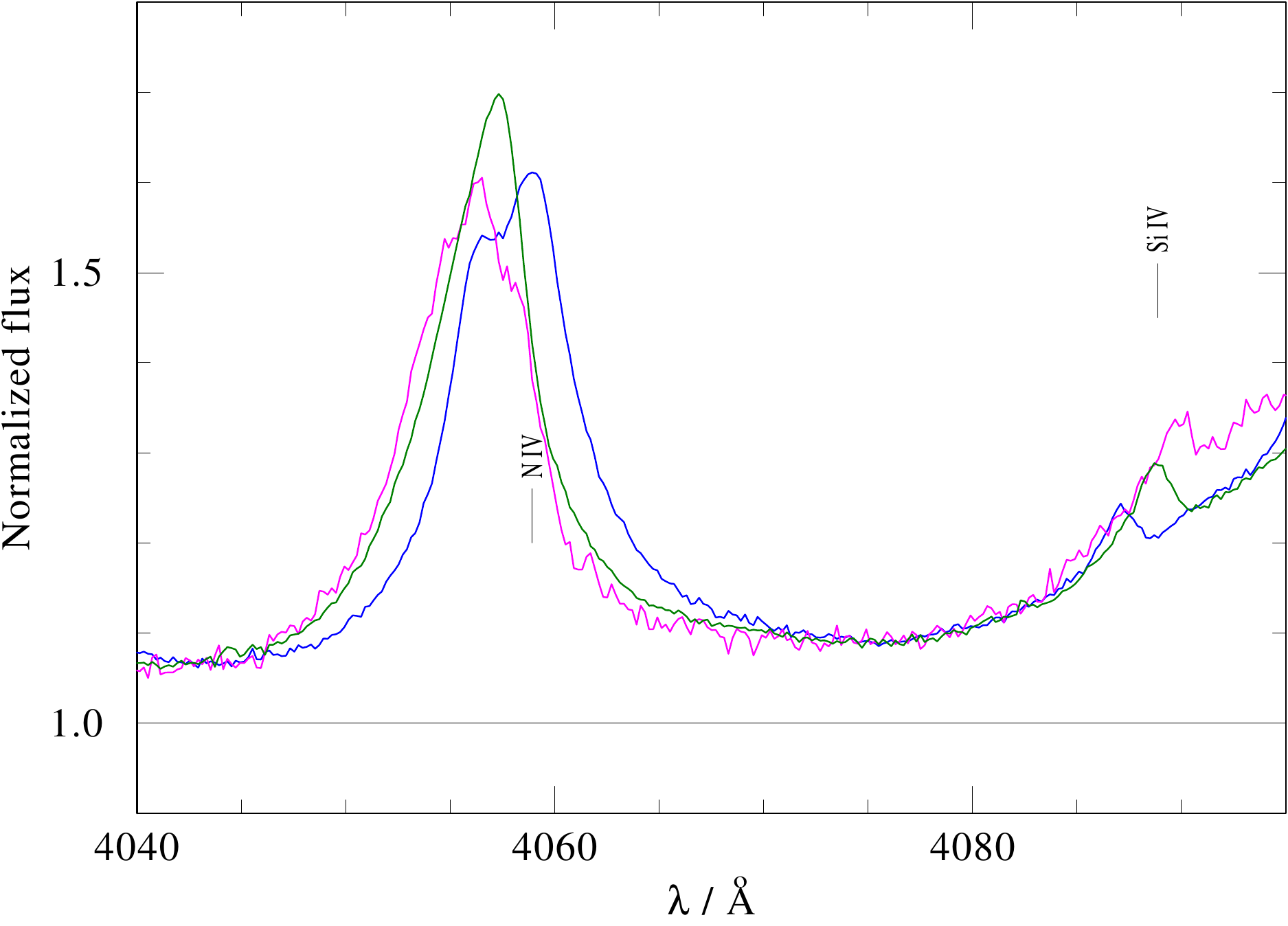}
  \caption{Zoom-in of Fig.\,\ref{fig:specoverview-flames} showing the N\,{\sc iv} and Si\,{\sc iv} lines 
moving in anti-phase. A FLAMES spectrum at $\phi \approx 0.5$ is also shown (green). The primary dominates the N\,{\sc iv} line, while the Si\,{\sc iv} lines 
originate solely in the secondary. The spectra are shifted by the systemic LMC velocity of 270\,\kms.
}
\label{fig:NIV-SiIV}
\end{figure}

Figure\, \ref{fig:specoverview-flames} shows two FLAMES spectra taken 
shortly before and after periastron ($\phi = 1$), where the RV differences are most extreme. 
It is readily seen that almost all spectral features shift in the same direction, 
i.e.\ they stem primarily from one component, the primary. However, a closer inspection reveals that the secondary contributes to the 
N\,{\sc iv}\,$\lambda 4058$ emission and to some absorption features seen on top of He\,II and Balmer lines. 
Most importantly, the Si\,{\sc iv} emission doublet at $\lambda \lambda 4089, 4116$ moves in pure antiphase to the majority of the available 
spectral features, implying that the Si\,{\sc iv} lines stem from the secondary alone. 
A zoom-in of Fig.\,\ref{fig:specoverview-flames} which focuses on the N\,{\sc iv} $\lambda 4058$ spectral region and a neighboring
member of the Si\,{\sc iv} doublet is shown in Fig.\,\ref{fig:NIV-SiIV}, where we also plot a spectrum taken close to apastron ($\phi \approx 0.5$). 

To measure the RVs of the components in the individual FLAMES spectra, we perform a 1D cross-correlation algorithm to different spectral features in all available spectra.
We tried two types of templates to cross-correlate with: a Gaussian, and the observations themselves. Both methods resulted in comparable values, although 
the Gaussians resulted in a worse fit quality, and are therefore omitted here.

We first used one of the FLAMES spectra as a template to cross-correlate with.
A calibration of the template to the restframe using rest wavelengths $\lambda_0$
is known to lead to systematic errors since the barycenter of emission lines rarely coincides with $\lambda_0$. Thus,
to calibrate the template, we cross-correlated specific features 
with two preliminary {\sc PoWR} models (see Sect.\,\ref{sec:specan}) calculated for the primary and secondary. 
We used the relatively symmetric and isolated  N\,{\sc iv}\,$\lambda 4058$ line and 
Si\,{\sc iv} doublet to calibrate the templates of the primary and secondary, respectively. 
Note that, while the calibration to the restframe depends to a certain extent on our {\sc PoWR} models, this only affects the 
systemic velocity $V_0$, and \emph{not} the remaining orbital parameters.

Once the two template spectra are calibrated, we cross-correlated them against the individual 
FLAMES spectra, identifying the RV with the maximum of the cross-correlation function. 
For the primary, we measured RV shifts using the lines N\,{\sc iv}\,$\lambda 4058$, He\,{\sc ii}\,$\lambda 4200$,  
He\,{\sc ii}\,$\lambda 4542$, and H$\gamma$. The secondary contributes to all these lines, but its contribution is significantly smaller due 
to its lower mass-loss rate, and, judging by the cross-correlation fits, 
we do not expect that it should influence the derived RVs. For the secondary, the 
Si\,{\sc iv} doublet offers the most reliable way to track the secondary's motion, 
although the weak N\,{\sc iii}\,$\lambda 4379$ line also shows a clear antiphase 
behavior, and was therefore measured for RVs. The Si\,{\sc iv} lines, which are 
blended with the H$\delta$ line, were rectified relatively to the contribution of the underlying 
H$\delta$ emission for a more accurate derivation of the RVs.

With the preliminary velocities 
at hand, we created "master templates`` for the primary and secondary by shifting the individual FLAMES spectra
to the rest frame and co-adding them. We then used these master templates to perform a new cross-correlation with the individual observations.
The newly derived RVs differed only slightly from the previous ones and generally show less scatter.
Errors on the RVs are estimated to be of the order of $10\,$\kms for the primary and $5\,$\kms for the secondary.

\begin{figure}[!htb]
\centering
  \includegraphics[width=\columnwidth]{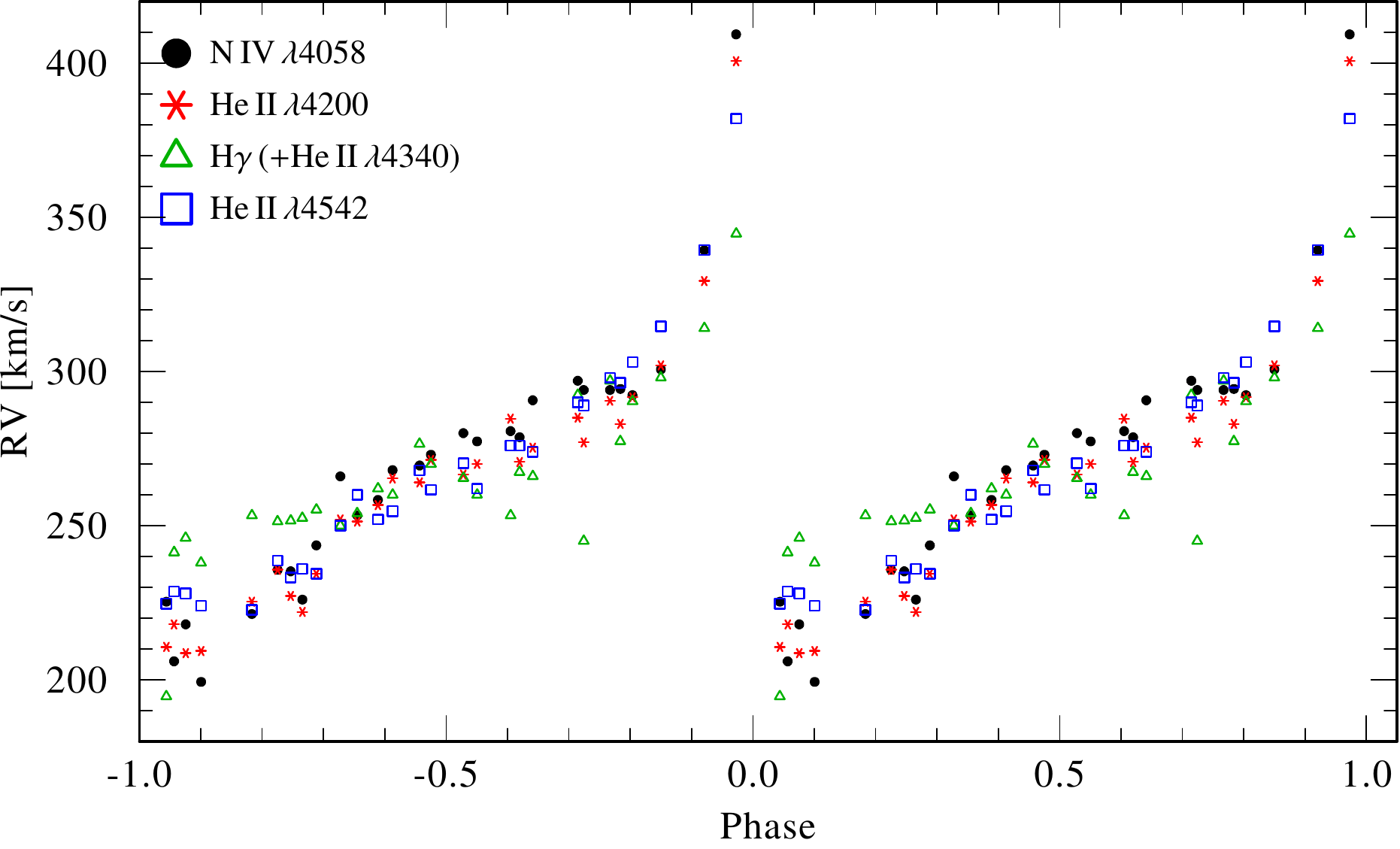}
  \caption{RVs for the primary component plotted over phase as measured for selected lines (see text).} 
\label{fig:radvelo-prim}
\end{figure} 

\begin{figure}[!htb]
\centering
  \includegraphics[width=\columnwidth]{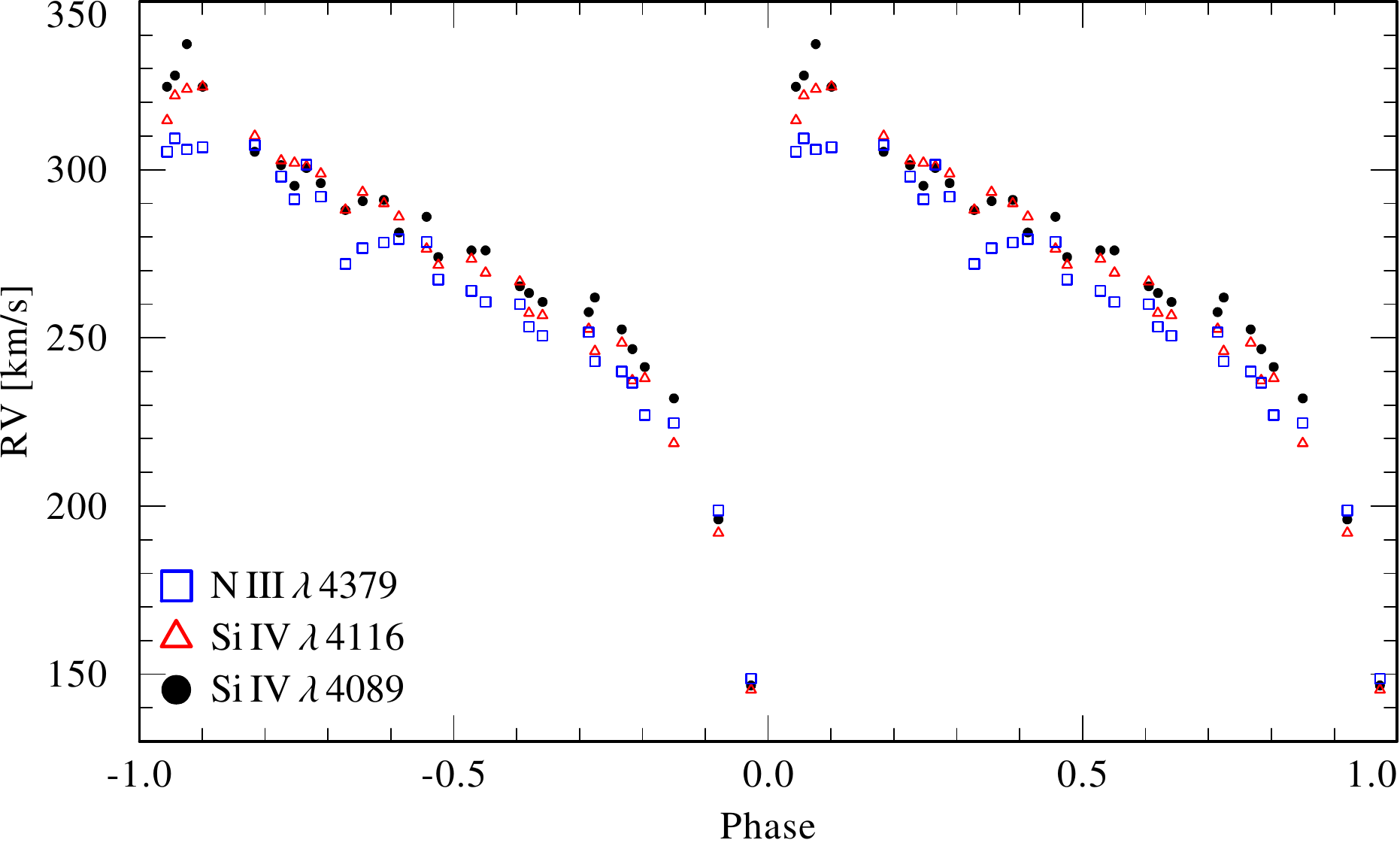}
  \caption{As Fig.\,\ref{fig:radvelo-prim}, but for the secondary.} 
\label{fig:radvelo-sec}
\end{figure}

The derived RVs for the primary and secondary as a function of phase are shown in Figs.\,\ref{fig:radvelo-prim} and 
\ref{fig:radvelo-sec}.
The phase is calculated relative to our final orbital solution (Table\,\ref{tab:orbpar}). 
The measured RVs for the N\,{\sc iv} (primary) and 
Si\,{\sc iv} lines (secondary) are listed in Table\,\ref{tab:RVs}.
The points are binned 
at intervals of $0.01$ on phase (note that often three FLAMES spectra were secured in a single observing night, 
cf.\ Table\,\ref{tab:RVs}).

From Fig.\,\ref{fig:radvelo-prim}, it is evident that N\,{\sc iv}\,$\lambda 4058$ and the He\,{\sc ii} $\lambda 4200$ and $\lambda 4542$ lines predict similar 
RVs, with N\,{\sc iv} showing a slightly larger velocity amplitude. In contrast, H$\gamma$ (+ He\,{\sc ii}\,$\lambda 4340$) shows significantly more scatter and 
a phase-dependent deviation from the other lines. 
As we will show in Sect.\,\ref{sec:wwc}, the reason is likely a contamination of H$\gamma$ by WWC, as well as its large formation radii.
The preliminary {\sc PoWR} model for the primary (see Sect.\,\ref{sec:specan})
suggests that the He\,{\sc ii} lines form a few stellar radii away from the stellar surface, 
as opposed to the N\,{\sc iv} line, which forms 
about $\approx 0.1$ stellar radii away. This implies that the N\,{\sc iv} line is more likely to represent the motion 
of the WR primary, and is therefore used for the orbital fitting.

As for the secondary, we use the Si\,{\sc iv} doublet. {\sc PoWR} models calculated 
for the secondary (see Sect.\,\ref{sec:specan}) suggest that it 
forms very close to the stellar surface ($\approx 0.05\,R_*$ above the stellar surface), 
and should be very 
reliable for measuring the secondary's RVs. 
The two Si\,{\sc iv} components agree well with each other and show a typical 
scatter of $\sigma \approx 7\,$\kms (Fig.\,\ref{fig:radvelo-sec}). For the final RVs of the secondary, we 
average the results of these two lines. 

In principle, we can derive orbital parameters using these velocities. 
However, the data only cover a few orbits and suffer from large gaps between them. 
We therefore combine 
old velocity measurements obtained by \citet{Moffat1989} and S2009 for the primary with our datasets
to assemble the longest possible time series, hence to obtain $P$ with the highest possible accuracy.
We note that the older velocities portray a significant systematic shift compared to the velocities derived here. This is mostly due to the different 
restframe calibration method used here. When performing the fitting, we therefore also fit for the  systematic shifts for both sets of velocities. 
The SB1 fitting procedure is performed  
through a Levenberg–Marquardt technique using the results of a 
Fourier analysis as a guess value for the initial period \citep{Gosset2001}.

From the SB1 fit, we derive the period $P_\text{orb} = 158.760\pm0.017$\,d, 
the epoch of periastron passage $T_0 \text{[MJD]} = 56022.4\pm0.8$, the RV amplitude of the primary $K_1 = 78\pm3\,$\kms, the eccentricity $e = 0.75\pm0.01$,
and the argument of periastron $\omega = 61\pm1^\circ$. 
Since the older velocities suffer from significantly larger errors, we do not adopt all orbital parameters derived, but only the period, which benefits significantly
from the $\approx 30$\,yr of coverage. We do not find evidence for apsidal motion in the system, which may, however, be a consequence 
of the fact that the new data are of much higher quality than previous ones.
In the next section, we analyze the polarimetric data simultaneously with the FLAMES data to better determine
the orbital parameters and the orbital inclination. 
Note that the assumption here is that the period change due to mass-loss from the system is negligible during these 
30 years. Since roughly $10^{-4.3}\,M_\odot$ are lost from the system each year (see Sect.\,\ref{sec:specan}), approx.\ 
$\Delta M_\text{tot} = 0.001\,M_\odot$ were lost within 30\,yr. The period change within 
30 years can be estimated via $P_\text{i} / P_\text{f} = \left(M_\text{tot, f} / M_\text{tot, i}\right)^2$ \citep{Vanbeveren1998}. 
Assuming $M_\text{tot} = 100\,M_\odot$ 
for an order-of-magnitude estimate, we obtain a difference in the period which smaller than our error, 
and thus negligible.

\section{Simultaneous polarimetry and RV fitting}
\label{sec:pol}

Fitting the polarimetric data simultaneously with the RV data enables us to lay tighter constraints 
on the orbital parameters. 
Furthermore, as opposed to an RV analysis, polarimetry can yield constraints on the inclination $i$. As the orbital masses 
scale as $M_\text{orb} \propto \sin^3 i$, knowing $i$ is crucial.

\renewcommand{\arraystretch}{1.2}

\begin{table}[!htb]

\caption{Derived orbital parameters}
\begin{center}
\begin{tabular}{l | c }
\hline      
Parameter                      & Value \\ 
\hline
$P_\text{orb}$ [days]                      & $158.760$              \\
$T_0$ [MJD]                     & $56022.6\pm0.2$                      \\
$K_1$ [\kms]                   &  $96\pm 3$                            \\ 
$K_2$ [\kms]                  &  $95\pm 4$                            \\
$e$                               &  $0.788\pm 0.007$            \\
$\omega [^\circ]$               &  $61\pm 7$             \\
$M_\text{orb, 1} \sin^3 i$ [$M_\odot$]     & $13.2\pm 1.9$                     \\
$M_\text{orb, 2} \sin^3 i$ [$M_\odot$]     & $13.4\pm 1.9$                     \\
$a_1 \sin i$ [$R_\odot$]     & $302\pm10$                       \\
$a_2 \sin i$ [$R_\odot$]     & $299\pm 10$                       \\
$V_0$ [\kms]                  & $270\pm5$                      \\ 
$\Omega$ [$^\circ$]                & $62\pm7$                      \\
$i$ [$^\circ$]                     & $39\pm6$                                  \\ 
$\tau_*$                        & $0.10 \pm 0.01$                 \\ 
$Q_0$                         & $-2.13 \pm 0.02$          \\
$U_0$                          & $0.58 \pm 0.02$         \\ 
$\gamma$                         & $0.87\pm0.07$             \\ 
$M_\text{orb, 1}$ [$M_\odot$]             & $53^{+40}_{-20}$                      \\
$M_\text{orb, 2}$ [$M_\odot$]             & $54^{+40}_{-20}$                      \\
$a_1 $ [$R_\odot$]           & $480^{+90}_{-65}$                         \\
$a_2 $ [$R_\odot$]           & $475^{+100}_{-70}$                         \\
\hline
\end{tabular}
\tablefoot{
Derived orbital parameters from a simultaneous fit of the FLAMES RVs and the polarimetry. The period is fixed to the value found from 
the SB1 fitting using all published RVs for the primary (see Sect.\,\ref{sec:RVmes}) 
}
\end{center}
\label{tab:orbpar}
\end{table}

The polarimetric analysis is based on ideas developed by \citet{Brown1978, Brown1982}, later corrected by \citet{Simmons1984}. 
A similar analysis for the system was performed by S2009.
As such, light emitted from a spherically-symmetric star is unpolarized. While Thomson scattering off free electrons in 
the stellar wind causes the photons to be partially linearly polarized, the total 
polarization measured in the starlight cancels out if its wind is spherically-symmetric. However, when the light of 
a star is scattered in the wind of its binary companion, the symmetry is broken, and some degree of polarization is expected. 
The degree of polarization depends on the amount and geometry of the scattering medium, which depends on the properties of the wind 
(e.g.,\ mass-loss) and on the orbital phase.

In our case, the dominant source of free electrons would clearly be the wind of the primary WR star (see also 
Sect.\,\ref{sec:specan}), although some of the primary's light may also be scattered in the wind of the secondary star. We first assume 
that only the wind of the primary contributes to the polarization, given its dominance over that of the secondary. 
We will later relax this assumption.
Following \citet{Robert1992}, the Stokes parameters $U(\phi)$ and $Q(\phi)$ can be written as the sum of the 
(constant) interstellar polarizations $U_0$, $Q_0$ and phase-dependent terms:

\begin{align}
\label{eq:UQpol}
\begin{split}
 U(\phi) &= U_0 + \Delta Q(\phi) \sin \Omega + \Delta U(\phi) \cos \Omega
\\ 
 Q(\phi) &= Q_0 + \Delta Q(\phi) \cos \Omega - \Delta U(\phi) \sin \Omega,
\end{split}
\end{align}
where $\Omega$ is the position angle of the ascending node, 
and $\Delta Q(\phi)$, $\Delta U(\phi)$ (in the case of spherically-symmetric winds) are given by

\begin{align}
\label{eq:UQdel}
\begin{split}
 \Delta U(\phi)  &= -2\tau_3(\phi) \cos i \sin (2\lambda(\phi))
\\ 
 \Delta Q(\phi) &= - \tau_3(\phi) \left[ \left(1 + \cos^2 i\right) \cos (2\lambda(\phi)) - \sin^2 i\right].
\end{split}
\end{align}
Here,  $\lambda(\phi)  = \nu(\phi) + \omega - \pi / 2$
is the longitude of the scattering source (primary) with respect to the illuminating source (secondary).  $\nu$ is the true anomaly, $\omega$ is the argument of periastron. 
$\tau_3$ 
is the effective optical depth of the scatterers \citep[see eqs.\,4,5 in][]{Robert1992} which scales with 
the (constant) total optical depth $\tau_*$ of the primary star \citep[see][]{Moffat1998b}. \citet{St-Louis1988} assumed 
that $\tau_3(\phi) = \tau_*\,(a / D(\phi))^\gamma$, where $D(\phi)$ is the separation between the companions, and 
$\gamma$ is a number of the order of unity. \citet{Brown1982} showed that 
$\gamma \approx 2$ in the case of a wind which is localized closely to the primary's stellar surface. However, this need not be the case 
for WR stars.

The free parameters involved in the polarimetry fitting are therefore $\Omega, i, \tau_*, Q_0, U_0$, and $\gamma$, as well as the orbital parameters 
$P, \omega,$ and $e$. One may generalize this model easily if both companions possess winds which can significantly contribute
to the total polarization. In this case, $\tau_*$ is the sum of optical depths of both stars, weighted with the relative light ratios 
\citep[see Eq.\ 2 in][]{Brown1982}. 
The formalism may be therefore implemented here, with the only consequence that $\tau_*$ relates to the mass-loss rates of both companions.

\begin{figure}[!htb]
\centering
  \includegraphics[width=0.9\columnwidth]{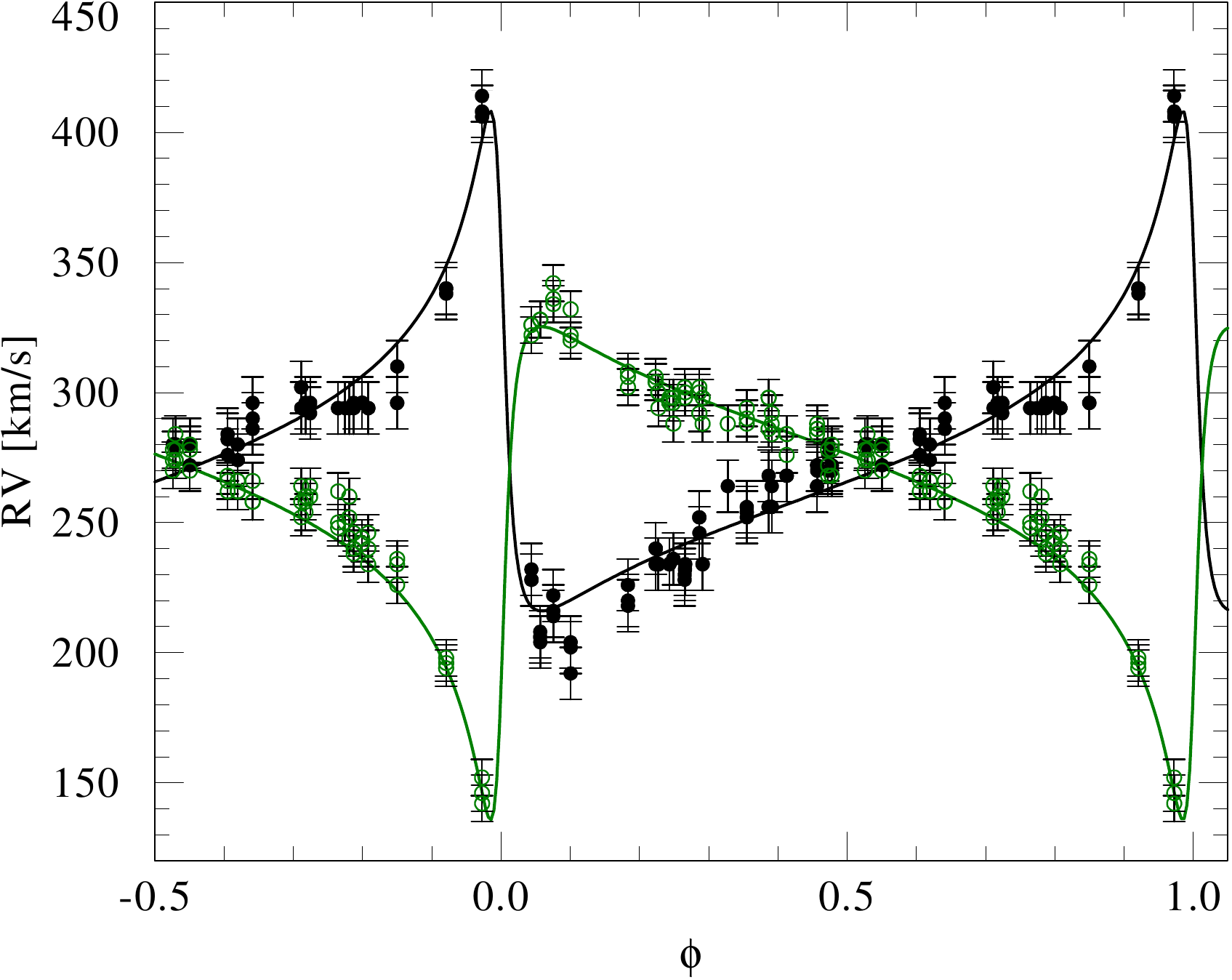}
  \caption{Orbital solution plotted against the measured RVs for the N\,{\sc iv}\,$\lambda 4058$ line (primary, black stars) and the averaged velocities of the 
  Si\,{\sc iv}\,$\lambda \lambda 4089, 4116$ doublet (secondary, green triangles).} 
\label{fig:RVfit}
\end{figure}

\begin{figure}[!htb]
\minipage{0.24\textwidth}
  \includegraphics[width=\linewidth]{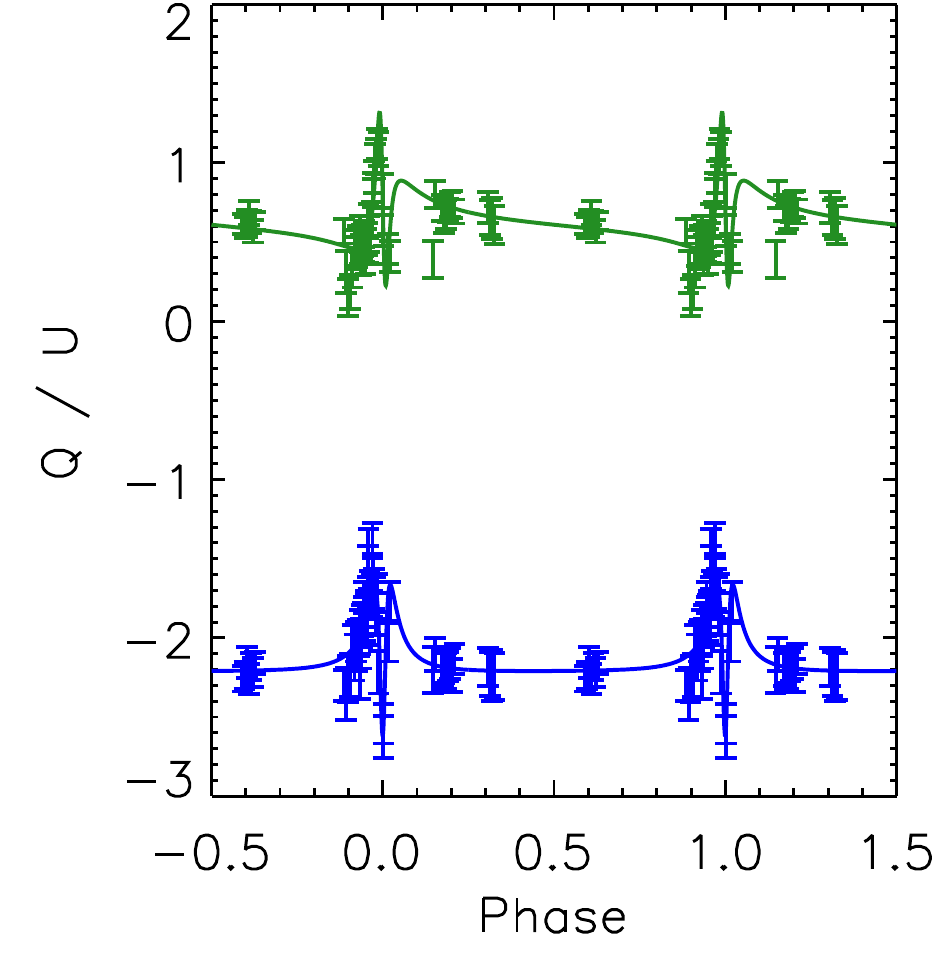}
\endminipage\hfill
\minipage{0.24\textwidth}
  \includegraphics[width=\linewidth]{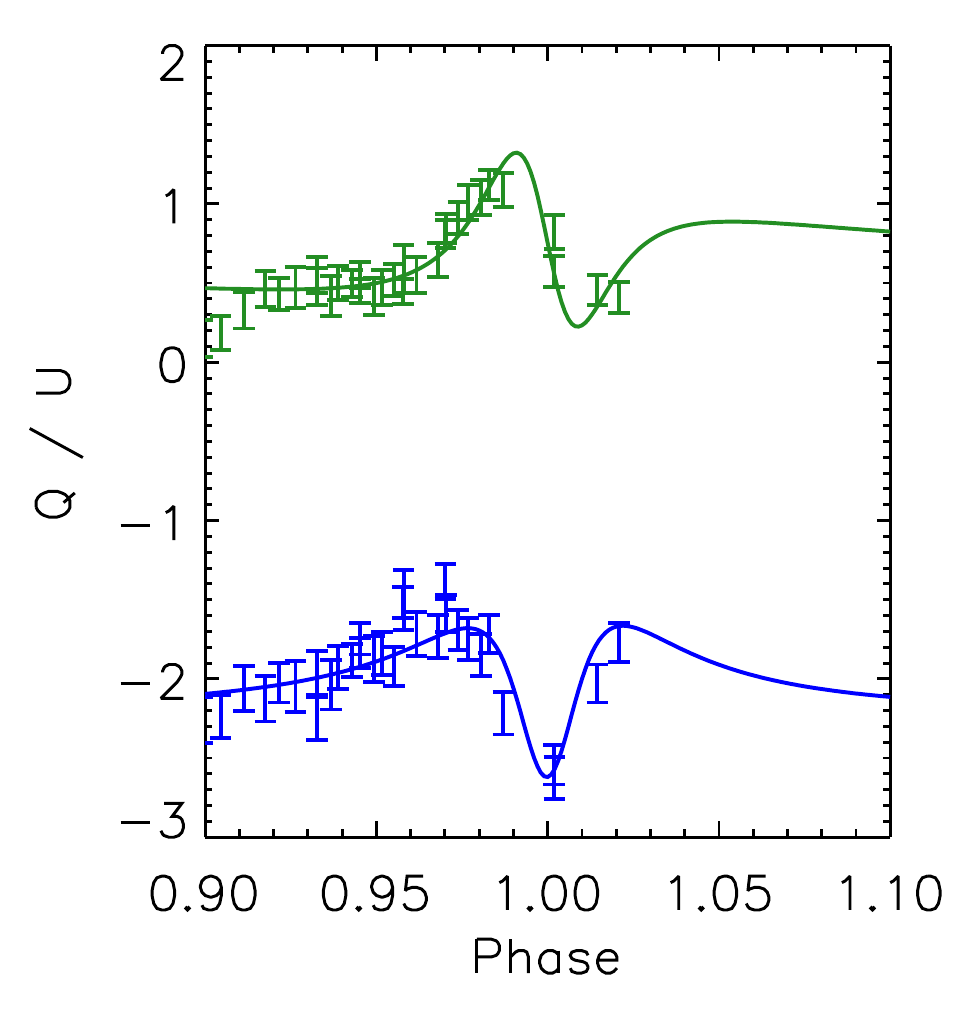}
\endminipage\hfill
\caption{Our polarimetric solution for Q (blue) and U (green) plotted against measured polarimetric data. 
The right panel is a zoom-in of the left panel around periastron passage}
\label{fig:polmods}
\end{figure}

The simultaneous fitting of the FLAMES RVs and the polarimetric data  is performed through a $\chi^2$ minimization algorithm,
with a relative weight given to the RV and polarimetric data chosen so 
that both types of data have a similar contribution to the total $\chi^2$. 
Best fit RV and Q/U curves are shown in Figs.\,\ref{fig:RVfit} and \ref{fig:polmods}.
During the fitting 
procedure, the period is fixed to the value inferred from the combined RV sample (see Sect.\,\ref{sec:RVmes}).
The corresponding best-fitting parameters are given in Table\,\ref{tab:orbpar}.


The inclination found in this study is very similar to that reported by S2009, which is not surprising 
given that we make use of the same polarimetric data.
We note that clumps in the wind can generally enhance the scattering and may therefore lead to an overestimation of the 
inclination.
The eccentricity is found to be larger, $e = 0.788\pm0.007$ 
as opposed to $e = 0.70\pm0.01$ found by S2009. 
This also affects the remaining orbital parameters (cf.\ Table\,5 in S2009). 
Most importantly, the orbital masses found here are much lower, $\approx 55^{+40}_{-20}\,M_\odot$ for each component 
compared to $M_1 \gtrsim 300$ and $M_2 \gtrsim 125$ which were inferred by S2009. The reason for this 
discrepancy is the improved derivation of $K_2$ in our study. 
While we cannot supply a definitive reason for the erroneous derivation of $K_2$ by S2009, we suggest that it may be 
related to the fact that the secondary exhibits emission features as well as a blue-shifted absorption. Furthermore, the spectra used in the latter 
study are of significantly lesser quality compared with the FLAMES spectra. 

The masses derived here are in much better agreement with the
brightness of the system, 
as discussed by S2009.
While the masses obtained here are close to the lower limit of what would be 
expected for these spectral types, as measured from eclipsing systems \citep[e.g.,][]{Rauw2004, Bonanos2004, Schnurr2008, Koenigsberger2014}, they are not inconsistent, 
especially considering the errors. Moreover, given the bias of the polarimetric fitting towards higher inclination angles 
\citep[e.g.,][]{Aspin1981, Wolinski1994}, the masses derived here are likely underestimated.

From Fig.\,\ref{fig:R136}, it is evident that the binary R145 is located outside the dense, massive cluster R\,136. 
The projected separation between the system and the cluster is only $\approx 20\,$pc, 
which, accounting for an age of  $\approx 2\,$Myr (cf.\ Table\,\ref{tab:inpar}), implies an average velocity of a mere $\approx 10\,$\kms.
The derived systemic velocity (cf.\, Table\,\ref{tab:orbpar}), which is comparable to the LMC mean velocity, would be consistent with 
a slow runaway ejected due to dynamical interactions within the cluster, as claimed for 
VFTS 682 \citep{Banerjee2012}. Alternatively, it may have formed in situ in the halo of the massive cluster R\,136.

\section{Spectral disentanglement}
\label{sec:disentangle}

Using the orbital parameters given in Table\,\ref{tab:orbpar} as an initial guess, we apply the disentanglement code {\sc Spectangular}
\citep{Sablowski2016} to the  
FLAMES spectra. 
The code performs the disentangling on a set of spectra in the wavelength domain rather than in the Fourier space \citep{Hadrava1995}.
Disentangling is coupled to an optimisation algorithm on the orbital parameters. Hence, it provides revised orbital parameters and the separated
component spectra.

It has been shown by \citet{Hadrava2009} that the disentangling 
can be successful as long as
the line-profile variability is small compared to a mean profile and the orbital motion. However, these conditions are hardly met by the system under study, 
both due to the intrinsic variability of WR stars (e.g.,\ due to clumping) as well as due to WWC. 
Indeed, we find that the orbital solution severely depends on the 
spectral domains used and the initial solution assumed.
We therefore adopt the orbital parameters obtained in Sect.\,\ref{sec:pol}.
In contrast, the resulting disentangled spectra are hardly influenced by the different solutions. We are therefore 
confident that the disentangled spectra obtained here represent the true spectra well, except
in cases where the lines are heavily contaminated by variability.

\begin{figure}[!htb]
\centering
  \includegraphics[width=\columnwidth]{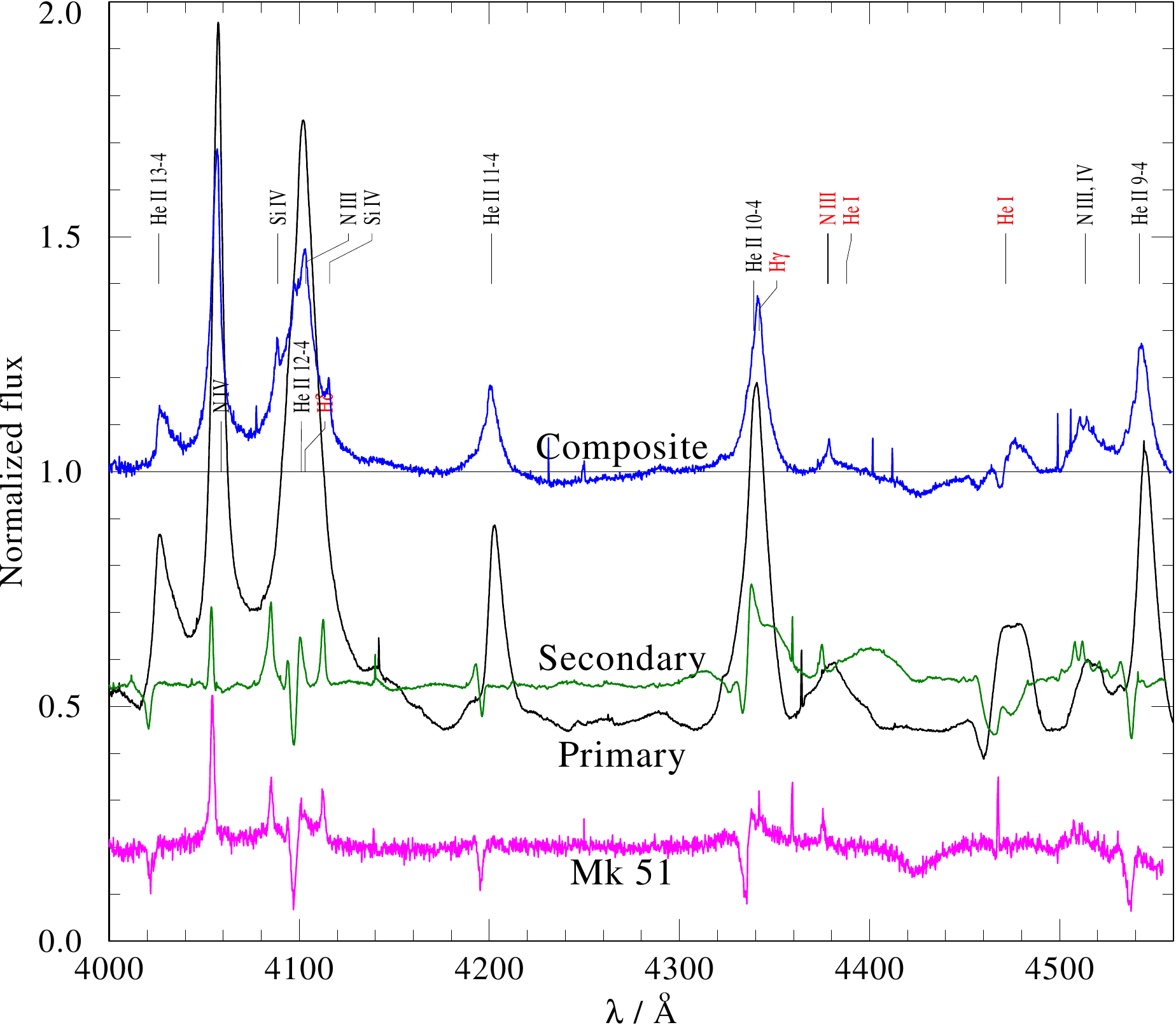}
  \caption{The disentangled normalized spectra of the primary (black) and secondary (green), shifted to their 
  relative light ratio. A composite FLAMES spectrum ($\phi \approx 0.5$) is shown for comparison 
  (blue), as well as an observed normalized spectrum of Mk 51 for comparison with the secondary, 
  shifted by -0.8 for clarity.} 
\label{fig:disentangled}
\end{figure}

With no eclipses in the system, it is impossible for the disentanglement procedure to provide the light ratio of the primary and secondary components. 
The adopted light ratio influences the strength of the lines in the disentangled, rectified spectra. 
Based on a calibration with equivalent widths (EWs) of putative single stars (see below), we
estimate the light ratio to be $F_{v,2} / F_{v,\text{tot}} = 0.55\pm0.10$, where $F_v$ is the 
visual flux in the Smith $v$ band. 
This ratio is assumed for the rectified disentangled spectra, 
shown in Fig.\,\ref{fig:disentangled}.

The disentangled spectrum of the primary is consistent with it being of WN6h type.
We compared the spectrum with two WN6h spectra in the LMC: \object{BAT99 30} and \object{BAT99 31}.  
The adopted light ratio results in EWs 
of the He\,{\sc ii} lines which agree well with the two latter objects,
but also results in an EW of the He\,{\sc i}\,$\lambda 4471$ line which is about three times
larger, suggesting a strong He\,{\sc i} excess in the system, likely originating in WWC (see Sect.\,\ref{sec:wwc}).

The spectrum of the secondary is suggestive of a so-called slash star \citep{Crowther2011}. Unfortunately, the FLAMES spectra 
do not include diagnostic lines which are important for the classification (e.g.,\ H$\beta$, N\,{\sc v}\,$\lambda \lambda 4603,4619$ and 
N\,{\sc iii}\,$\lambda \lambda \lambda 4634,4640,4642$). Moreover, some lines, marked in red in Fig.\,\ref{fig:disentangled}, are strongly affected by WWC 
(see Sect.\,\ref{sec:wwc}). 
This is especially significant for the secondary's spectrum, which generally shows weaker features compared to the primary. Especially the shape and strength 
of the lines in the range $4000-4200\,\AA$ is suggestive of an O3.5~If*/WN7 star, for which the star \object{Melnick 51} (\object{Mk 51}) is a prototype. 
For comparison, we plot in Fig.\,\ref{fig:disentangled} an observed normalized FLAMES spectrum of Mk 51 (P.\ Crowther, priv.\ com.).
The Si\,{\sc iv} doublet is stronger in R145 than observed for MK51, and the N\,{\sc iv} line weaker. 
However, the derived model 
and parameters for the secondary (see Sect.\,\ref{sec:specan})
are suggestive of the 
spectral class O3.5~If*/WN7. We therefore adopt this spectral class in this study. 
The light ratio adopted here is also chosen so that the EWs of the majority of 
the Balmer and He\,{\sc ii} lines agree with this spectral type (see Fig.\,\ref{fig:disentangled}).

\section{Spectral analysis}
\label{sec:specan}

The disentangled spectra, together with the high-quality XSHOOTER spectrum and the complementary UV and photometric data, 
enable us to perform a spectral analysis of both components. 
The spectral analysis is performed with the Potsdam Wolf-Rayet\footnote{{\sc PoWR} models of Wolf-Rayet stars can be downloaded at
http://www.astro.physik.uni-potsdam.de/PoWR.html} ({\sc PoWR}) model atmosphere code, 
applicable to any hot star \citep[e.g.,][]{Shenar2015, Todt2015, Angel2016}. 
The code iteratively solves the co-moving frame, non-local thermodynamic equillibrium (non-LTE) radiative transfer and the statistical balance equations in spherical symmetry under the constraint of energy conservation, yielding
the occupation numbers in the photosphere and wind. By comparing the output synthetic spectra to observed spectra, fundamental stellar parameters are derived.
A detailed description of the assumptions and methods used
in the code is given by \citet{Graefener2002} and \citet{Hamann2004}. Only essentials 
are given here.

A {\sc PoWR} model is defined by four fundamental stellar parameters: the effective temperature $T_*$, the surface gravity $g_*$, the stellar luminosity $L$, and the mass-loss rate $\dot{M}$. 
The effective temperature $T_*$ is given relative to the stellar radius $R_*$, so that $L = 4\,\pi\,\sigma\,R_*^2\,T_*^4$. 
$R_*$ is defined at the model's inner boundary, fixed at mean Rosseland optical depth of 
$\tau_\text{Ross} = 20$ \citep{Hamann2006}. The outer boundary is set to $R_\text{max} = 1000\,R_*$.
The gravity $g_*$ relates to the radius $R_*$ and mass $M_*$  via the usual definition: $g_* = g(R_*) = G\,M_* R_*^{-2}$. 
We cannot derive $g_*$ here because of the negligible effect it has on the wind-dominated spectra, and fix it to the value 
implied from the orbital mass.

\renewcommand{\arraystretch}{1.2}
\setlength{\tabcolsep}{2mm}

\begin{table}[!htb]
\scriptsize
\small
\caption{Derived physical parameters for R145.}
\label{tab:specan}
\begin{center}
\begin{tabular}{l | c  c}
\hline      
Parameter & Primary & Secondary \\ 
\hline
Spectral type        & WN6h & O3.5~If*/WN7 \\
$T_*$ [K]        & $50000\pm3000$ & $43000\pm3000$ \\
$\log L$ [$L_\odot$]  &  $6.35\pm0.15$   &  $6.33\pm0.15$ \\ 
$\log R_\text{t}$ [$R_\odot$]  & $1.05\pm0.05$   &  $1.50\pm0.15$ \\
$v_\infty$ [\kms] &    $1200\pm200$  & $1000\pm200$ \\
$R_*$ [$R_\odot$]  & $20^{+6}_{-5}$  & $26^{+9}_{-7}$ \\ 
$D$ [$R_\odot$] & $10\pm0.3\,$dex & $10\pm0.3\,$dex \\
$\log \dot{M}$ [$M_\odot\,{\rm yr}^{-1}$] & $-4.45\pm0.15$ & $-4.9\pm0.3$  \\
$v \sin i$ [\kms] & $< 200$ & $< 150$ \\
$v_\text{rot}$ [\kms] & $< 350$ & $< 270$ \\
$X_\text{H}$ (mass fraction) &  $0.4\pm0.1$ & $0.5\pm0.2$  \\
$X_\text{C} / 10^{-4}$ (mass fraction) & $0.8\pm0.3$ & $0.7\pm0.4$ \\
$X_\text{N} / 10^{-3}$ (mass fraction) & $8\pm4$ & $8\pm4$ \\
$X_\text{O} / 10^{-4}$ (mass fraction) & $\lesssim 1$  & $\lesssim 1$ \\ 
$X_\text{Si} / 10^{-4}$ (mass fraction) & $7$ (fixed) & $7\pm3$ \\ 
$M_\text{v}$[mag] & $-7.21\pm0.25$ & $-7.43\pm0.25$ \\
$M_\text{V}$[mag] & $-7.15\pm0.25$ & $-7.30\pm0.25$ \\
$M_\text{MLR,hom}[M_\odot]$ & $101^{+40}_{-30}$ & $109^{+60}_{-40}$ \\
$M_\text{MLR,He-b}[M_\odot]$ & $55^{+15}_{-12}$ & $54^{+15}_{-12}$ \\
$E_{B-V}$ [mag] & \multicolumn{2}{c}{$0.34\pm0.01$} \\
$A_{V}$ [mag] & \multicolumn{2}{c}{$1.4\pm0.2$} \\
\hline
\end{tabular}
\end{center}
\end{table}

\begin{figure*}[]
\centering
  \includegraphics[width=0.85\textwidth]{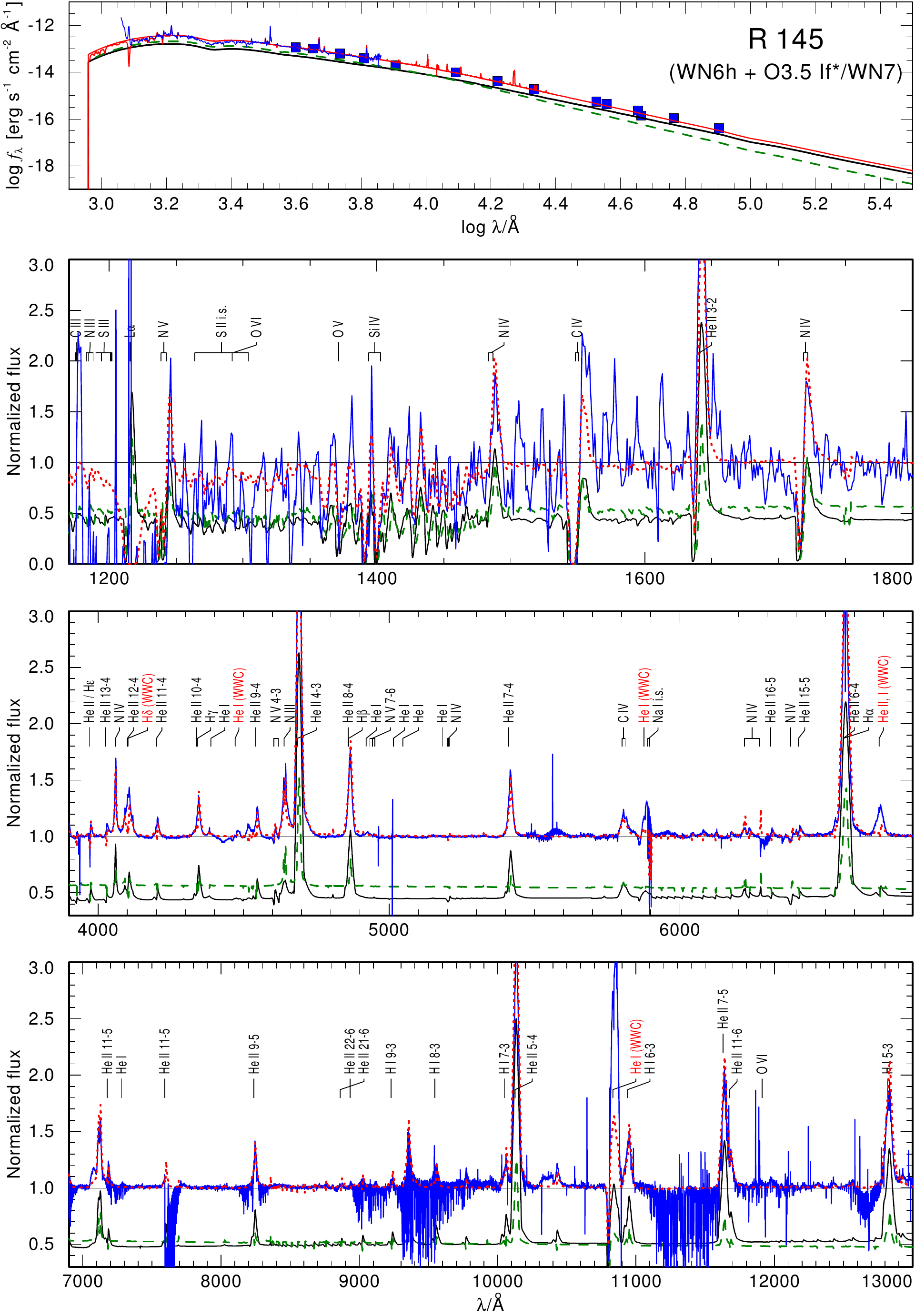}
  \caption{Comparison between observed (blue squares and lines) SED (upper panel) and the normalized IUE and XSHOOTER spectra (lower panel) and the synthetic composite spectrum (red dotted line). 
  The composite spectrum is the sum of the primary (black solid line) and secondary (green dashed line). The observed and modelled spectra in the UV  
  are binned at $1\AA$ for clarity. Lines which are strongly affected by WWC are marked with red idents.}
\label{fig:specan}
\end{figure*} 

The chemical abundances of the elements included in the calculation are prespecified.
Here, we include H, He, C, N, O, Si, and the iron group elements dominated 
by Fe. The mass fractions $X_\text{H}$, $X_\text{C}$, and $X_\text{N}$, and $X_\text{Si}$ are derived in this work. Based on studies by 
\citet{Korn2000} and \citet{Trundle2007}, we set $X_\text{Fe} = 7\cdot10^{-4}$. Lacking any signatures 
associated with oxygen, we fix $X_\text{O} = 5\cdot10^{-5}$ for both components. Values larger than $10^{-4}$ lead to spectral 
features which are not observed. 

Hydrostatic equilibrium is assumed in the subsonic 
velocity regime \citep{Sander2015}, from which the density and velocity profiles follow, while a $\beta$-law \citep{CAK1975} with $\beta = 1$ 
\citep[e.g.,][]{Schnurr2008}
is assumed for the supersonic regime, defined by 
the $\beta$ exponent and the terminal velocity $v_\infty$. Optically thin clumps are accounted for using the microclumping approach 
\citep{Hillier1984, Hamann1998},
where the population numbers are calculated in clumps which are a factor of $D$ denser 
than the equivalent smooth wind ($D = 1 / f$, where $f$ is the filling factor). 
Because optical WR spectra  are dominated by recombination lines, whose strengths increase with $\dot{M} \sqrt{D}$, it is customary 
to parametrize their models using the so-called transformed radius \citep{Schmutz1989},

\begin{equation}
 R_\text{t} = R_* \left[ \frac{v_\infty}{2500\,{\rm km}\,{\rm s}^{-1}\,}  \middle/  
 \frac{\dot{M} \sqrt{D}}{10^{-4}\,M_\odot\,{\rm yr}^{-1}}  \right]^{2/3},
\label{eq:Rt}
\end{equation}
defined so that EWs of recombination lines of models  
with given abundances, $T_*$, and $R_\text{t}$ are 
approximately preserved, independently 
of $L$, $\dot{M}$, $D$, and $v_\infty$.

The effective temperature of the primary is derived mainly based on the ionisation balance of N\,{\sc iii}, N\,{\sc iv}, and N\,{\sc v} lines. For the secondary, the weakness of 
associated He\,{\sc i} lines, as well as the presence of a strong N\,{\sc iii} component and a weak N\,{\sc iv} component constrain $T_*$.
Once the temperatures and light ratio (see Sect.\,\ref{sec:disentangle}) are constrained, mass-loss rates (or transformed radii) can be determined. For the primary, this is
straightforward, while for the secondary, this can 
only be done approximately. The terminal velocity $v_\infty$ is determined primarily from P-Cygni lines in the UV.
Clumping factors are determined using electron scattering wings, primarily of He\,{\sc ii}\,$\lambda 4686$. Hydrogen content is derived based on the balance 
of the Balmer series (He\,{\sc ii} $+$ H) to pure He\,{\sc ii} lines. The remaining abundances are derived from the overall strengths of their associated lines.

The luminosity 
and reddening follow from a simultaneous fit to available photometry, adopting a distance of 50\,kpc \citep{Pietrzynski2013}. 
We use U photometry from \citet{Parker1992}, BVRI photometry from \citet{Zacharias2012},  JHK and IRAC photometry from the compilation of
\citet{Bonanos2010}, and WISE photometry from \citet{Cutri2013}. 
The reddening is 
modelled using the reddening law published by \citet{Howarth1983}. In the latter, we find $R_\text{V} = 4.0\pm0.5$ is 
most consistent in reproducing the complete photometry, comparable to other stars in 30 Dorados \citep{Maiz2014}, 
and we therefore fix $R_\text{V} = 4$ and fit for $E_\text{B - V}$. \cite{Maiz2014} derived new laws for the 30 Dor region, but since the difference 
between these laws and older ones are negligible in the reddening regime involved here (see figures 11 and 12 in the latter paper) - 
especially for the purpose of this study -  these new laws are not implemented here.

The nitrogen abundance is found to be about a factor of two larger in both components 
compared to the typical LMC values \citep[cf.][]{Hainich2014}, mostly due to the strong N\,{\sc iii} doublet at $\approx 4640\,\AA$. 
However, this enhancement may be insignificant given the errors.
Furthermore, 
to reproduce the Si\,{\sc iv} doublet originating in the secondary, 
it is necessary to set $X_\text{Si}$ to an abundance comparable to the Galactic one ($\approx$ two times larger than typical 
LMC abundance, cf.\ \citealt{Trundle2007}). Since $X_\text{Si}$ is not expected to change 
throughout the stellar evolution, we assume that silicon was initially overabundant, and fix the same value for the primary.
Unfortunately, the poor quality of the UV data does not 
enable us to determine the abundance of the iron group elements. Because of the relatively large associated errors, we refrain 
from interpreting this apparent overabundance.

A comparison of the best-fitting models to the observed spectral energy distribution (SED) and normalized spectra is shown in Fig.\,\ref{fig:specan}. 
Note that the composite spectrum strongly underpredicts low-energy transitions 
such as He\,{\sc i} lines. We will show in Sect.\,\ref{sec:wwc} that these lines are expected to be strongly contaminated by WWC.
The derived stellar parameters are listed in Table\,\ref{tab:specan}, where we also give the Smith and Johnson absolute magnitudes $M_\text{v}$ and $M_\text{V}$, 
as well as the total extinction $A_\text{V}$. 
We also give upper limits derived for the projected and actual rotation velocity $v \sin i$ and $v_\text{rot}$ 
for both components, as derived by comparing lines formed close to the stellar surface (N\,{\sc iv}\,$\lambda 4058$, Si\,{\sc iv}\ doublet) 
to synthetic spectra which account for rotation in an expanding atmosphere, assuming co-rotation up to $\tau_\text{Ross} = 2/3$ and 
angular momentum conservation beyond \citep[cf.][]{Shenar2014}. Given the low inclination angle, these only lay weak constraints on the actual 
rotation velocities $v_\text{rot}$ of the stars.
Errors are estimated from the sensitivity of the fit quality to 
variations of stellar parameters, or via error propagation. 

Table\,\ref{tab:specan} further gives stellar masses which are based on mass-luminosity relations calculated by \citet{Graefener2011} for homogeneous stars. 
The relations depend on $L$ and $X_\text{H}$ alone. $M_\text{MLR, hom}$ assumes the derived value of $X_\text{H}$ in the core, i.e.\ a 
homogeneous star. $M_\text{MLR, He-b}$ assumes $X_\text{H} = 0$, i.e.\ the relation for pure He stars, which is  
a good approximation if the hydrogen rich envelope is of negligible to moderate mass (see discussion by \citealt{Graefener2011}). 
If indeed $M_\text{orb} \approx 55\,M_\odot \approx M_\text{MLR, He-b}$, as is implied from Tables\,\ref{tab:orbpar} and \ref{tab:specan}, 
the stars are likely already core He-burning. However, 
this is very unlikely to be true for the secondary given its spectral type and luminosity. 
Rather, the orbital masses are likely underestimated due to an overestimated 
inclination \citep[e.g.,][]{Aspin1981}, and may in fact be more similar to $M \approx 80-90\,M_\odot$, which is consistent with the upper boundary of our 
errors (see further discussion in Sect.\,\ref{sec:disc}).

Within errors, the derived physical parameters are in good agreement with the spectral types of the primary and secondary \citep[cf.][]{Crowther2011, Hainich2014}. 
\citet{Bestenlehner2014} and \citet{Hainich2014} both analyzed R145 assuming a single component, which explains why they derive a luminosity in 
excess of $\log L = 6.5\,[L_\odot]$, about $0.2-0.3\,$dex higher than found here for the primary. \citet{Hainich2014} found a comparable effective temperature 
to that found for the primary in our study, while \citet{Bestenlehner2014} found a significantly lower temperature of $40\,$kK (comparable to the secondary), 
which is a consequence of attributing strong features stemming partially from the secondary (e.g.,\ strong N\,{\sc iii} lines) to the primary. 
Similarly, the mass-loss rates derived here are different than in the previous studies because they did not account for line dilution and adopted 
wrong luminosities. Given the careful binary analysis performed here, we are inclined to believe that our results represent 
the system much more accurately 
than previous studies.

\section{Variability and wind-wind collision}
\label{sec:wwc}

Our results from the previous sections imply that both binary components in R145 have significant stellar winds. 
In this case, it is expected that a cone-shaped wind-wind collision (WWC) zone would form, its tip situated along the line connecting the centers of both 
stars at the point where 
the dynamical pressures of the two outflows equalize \citep{Stevens1992, Moffat1998}. The temperatures at the immediate vicinity of the collision zone 
can reach a few $10^7\,$K. The plasma rapidly cools and emits radiation as it streams outwards along the cone.

Observationally, the emission takes on two forms. On the one hand, WWC leads to excess emission which can be seen photometrically,
either in X-rays \citep{Cherepashchuk1976, Corcoran1996} or 
in non-thermal radio, 
infra-red and optical 
\citep[e.g.,][]{Williams1997}. If the WWC occurs after the winds have reached their terminal velocities,  
the overall strength of the emission 
is expected to reach a maximum at periastron. In the adiabatic case, an inverse proportion 
with the separation D between the stars is predicted \citep{Usov1992}.
A sharper scaling ($\propto D^{-n}$, $n > 1$) is expected in highly radiative cases.
On the other hand, WWC emission can also be spotted spectroscopically on top of prominent emission lines in the optical, as the plasma cools off
via recombination \citep[e.g.,][]{Rauw1999, Hill2000, Sana2001}. A spectroscopic 
analysis of the excess emission arising from the WWC zone can place strong constraints on the kinematics and inclination of the system \citep{Luehrs1997}.

\begin{figure}[!htb]
\centering
  \includegraphics[width=\columnwidth]{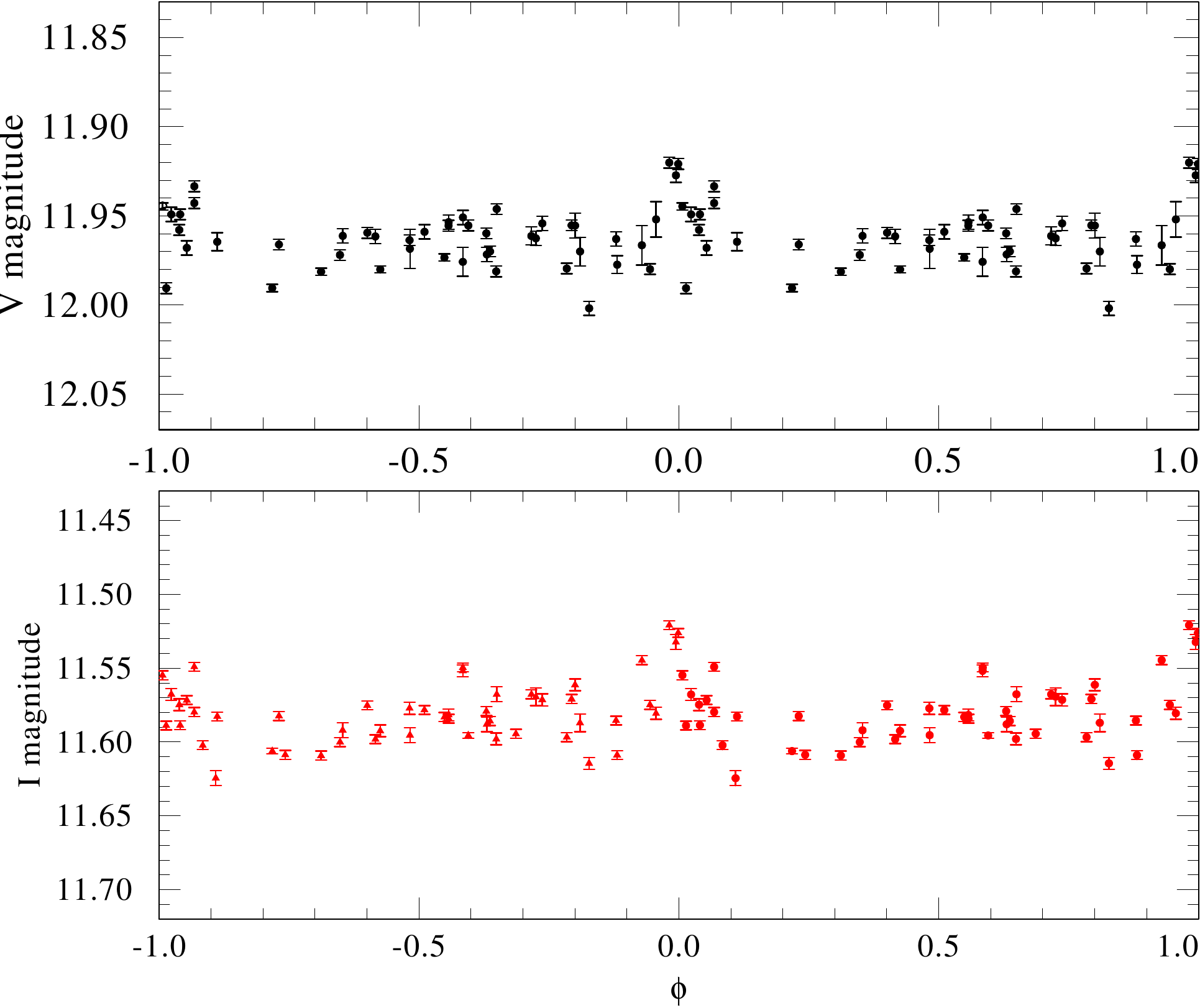}
  \caption{Phased OGLE-III V- and I-band light-curve of R145.}
\label{fig:lightcurve}
\end{figure}

In Fig.\,\ref{fig:lightcurve}, we show V- and I-band light-curve of R145 (K. Ulaczyk, private communication) obtained with the 
OGLE-III shallow survey \citep{Ulaczyk2012}, phased with the ephemeris in Table\,\ref{tab:orbpar}. One can 
see a clear emission excess of $\approx 5\%$ during periastron passage. Possible mechanisms which could 
cause a phase-dependent variability include wind eclipses, ellipsoidal deformations \citep[e.g.,][]{Soszynski2004}, 
and WWC. Wind eclipses are expected to cause a dip as the components align along the line of sight 
(phases $0.01$ and $0.82$), and while the outliers seen in Fig.\,\ref{fig:lightcurve} around these phases could indicate a wind eclipse, the data points are too 
sparse to tell. Ellipsoidal deformations could play a role, although it is unclear whether they are expected to be 
important for spectra which are wind-dominated. However, the emission excess seen during periastron is most easily explained 
by WWC. Given the sparseness of the data, however, we refrain from modelling the light-curve.


\begin{figure}[!htb]
\centering
  \includegraphics[width=\columnwidth]{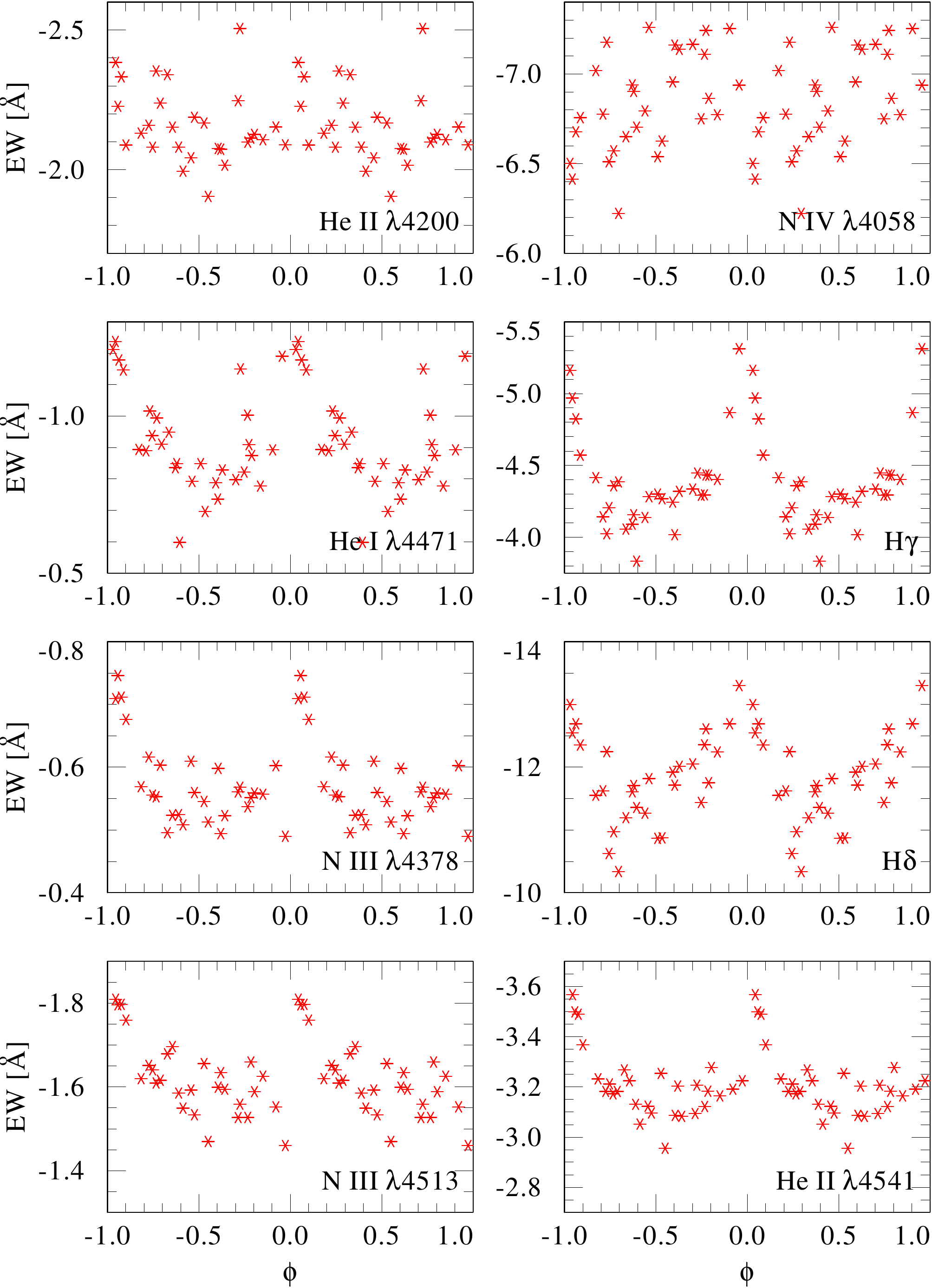}
  \caption{EWs as a function of orbital phase $\phi$ measured for selected lines in the FLAMES spectra, binned at $\Delta \phi = 0.01$ intervals.} 
\label{fig:WWC-EWs}
\end{figure}

\begin{figure}[!htb]
\centering
  \includegraphics[width=\columnwidth]{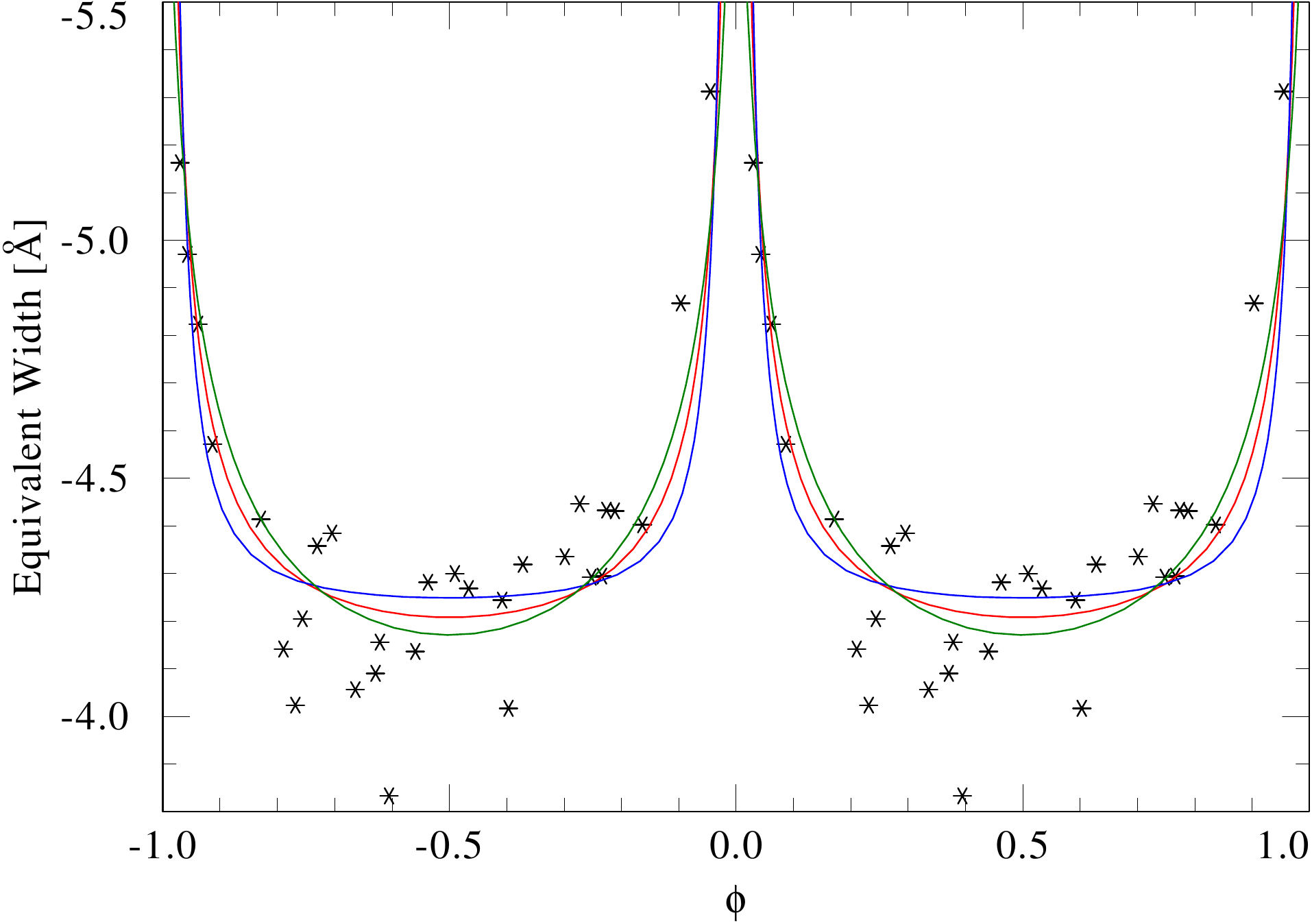}
  \caption{Fits of the form $A_1 + B_1\,D^{-2}$ (blue curve), $A_2 + B_2\,D^{-1}$ (red curve), and $A_3 + B_3\,D^{-\alpha}$ (green curve) to the data points
   describing the EW as a function of phase $\phi$ for the line H$\gamma$. } 
\label{fig:WWC-fit}
\end{figure}

\begin{figure}[!htb]
\centering
  \includegraphics[width=0.8\columnwidth, angle=-90]{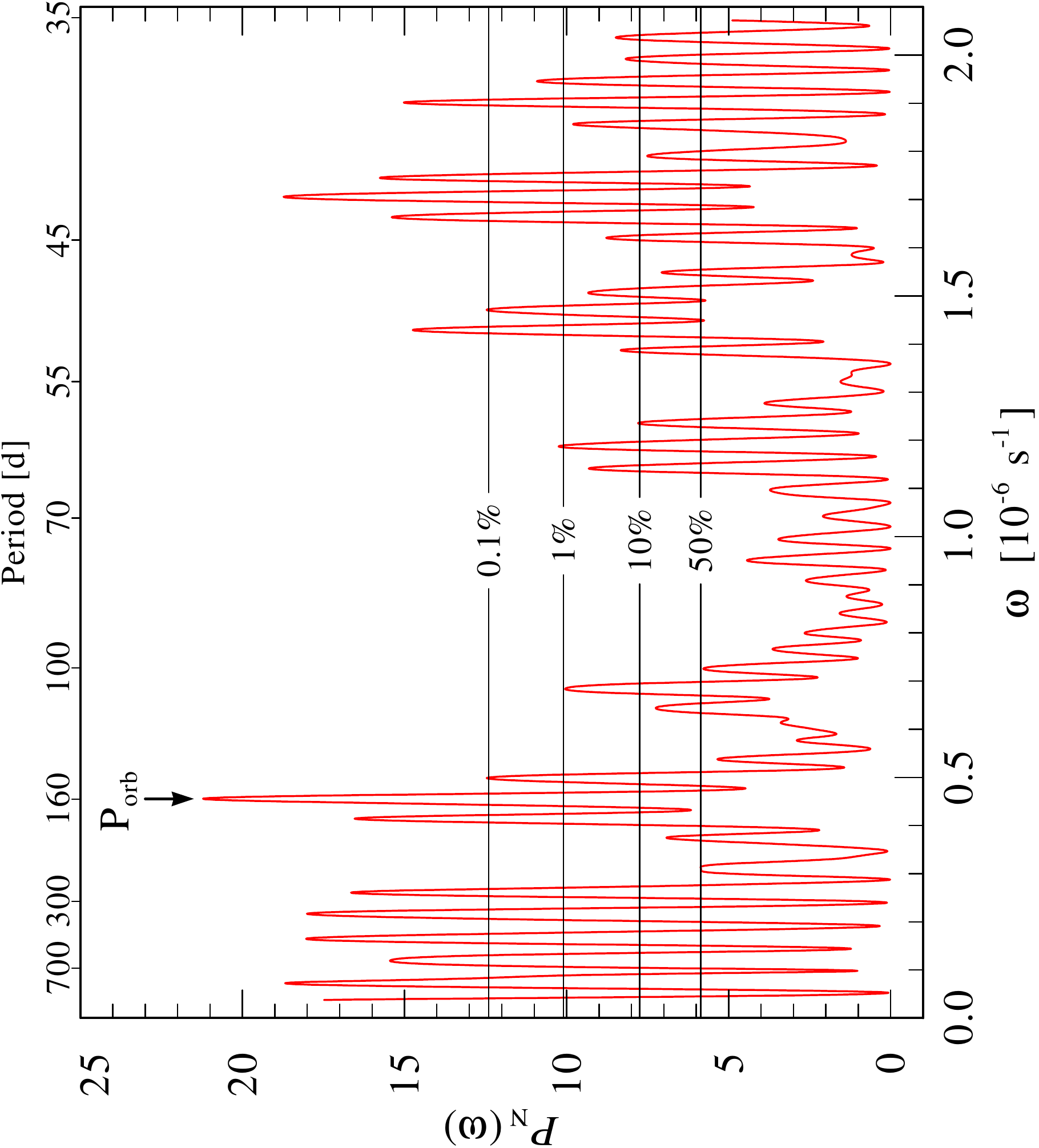}
\caption{Periodogram of the EWs of the H$\delta$ line.
The periodogram was calculated from $\omega = 2\pi/T$ to $\omega = \pi N_0/T$ with a spacing $0.1/T$, where $N_0$ is the number of data points and $T$ the total
time of the observation. Various false-alarm probability levels are marked.
}
\label{fig:WWC-scargle}
\end{figure} 

To study the spectroscopic variability, we calculated the EWs of several lines in all 110 FLAMES spectra 
to check for periods present in the dataset. 
Figure\,\ref{fig:WWC-EWs} shows the EWs plotted versus phase, binned on intervals of $\Delta \phi = 0.01$.
It is evident that lines associated with ``cooler'' ions such as He\,{\sc i}, N\,{\sc iii}, and Balmer lines show a clear increase 
of the emission near periastron ($\phi = 0$). The largest increase in flux (factor of two)
is seen at the He\,{\sc i}\,$\lambda 4471$ transition, followed by an increase of $\approx 40\%$ for $H\gamma$. This is significantly more 
that observed for the continuum (see Fig.\,\ref{fig:lightcurve}).
This behavior is not seen at all in the He\,{\sc ii}\,$\lambda 4200$ and N\,{\sc iv}\,$\lambda 4058$ lines, but is seen in the 
He\,{\sc ii}\,$\lambda 4541$ line, possibly because it is blended with an N\,{\sc iii} component.

In Fig.\,\ref{fig:WWC-fit}, we plot 
the same data points as in Fig.\,\ref{fig:WWC-EWs} for H$\gamma$, but include three curves which correspond to functions of the 
form $A_1 + B_1\,D^{-1}(\phi)$ , $A_2 + B_2\,D^{-2}(\phi)$, and $A_3 + B_3\,D^{-\alpha}(\phi)$ with 
the constants $A_\text{i}$, $B_\text{i}$, and $\alpha$ chosen to minimize the sum of the squared differences $\chi^2$. 
When leaving the exponent $\alpha$ as a free parameter, we obtain $\alpha \approx 0.25$. A similar test for N\,{\sc iii}\,$\lambda 4378$ line 
results in $\alpha \approx 1$, while for the He\,{\sc ii}\,$\lambda 4541$ line, we obtain $\alpha \approx 1.2$. Given the intrinsic scatter in the EWs 
and the poor coverage during periastron, we cannot exclude the $1/D$ adiabatic 
dependence predicted by \citet{Usov1992}.

We checked for the presence of periodic signals on the EWs of the lines shown in Fig.\,\ref{fig:WWC-EWs}. In most cases, we find significant 
detections of periods which agree with the orbital period. The remaining periods are found to be insignificant. In Fig.\,\ref{fig:WWC-scargle}, 
a periodogram \citep{Scargle1982, Horne1986}  is shown as an example. 
The most prominent peak appears for a period of
$158.9 \pm 0.8$ days, in very good agreement with the orbital period found (cf.\ Table\,\ref{tab:orbpar}). The occurrence of further apparently significant peaks 
is caused primarily by spectral leakage due to the 
unevenly spaced data \citep{Horne1986}, as we confirmed by subtracting the main signal and constructing a second periodogram. We find a marginal detection 
of a further period of $P_2 \approx 21\,$d, which may be related to stochastic variability in the system, but could also be spurious.

\begin{figure*}[!htb]
\minipage{0.32\textwidth}
  \includegraphics[width=\linewidth]{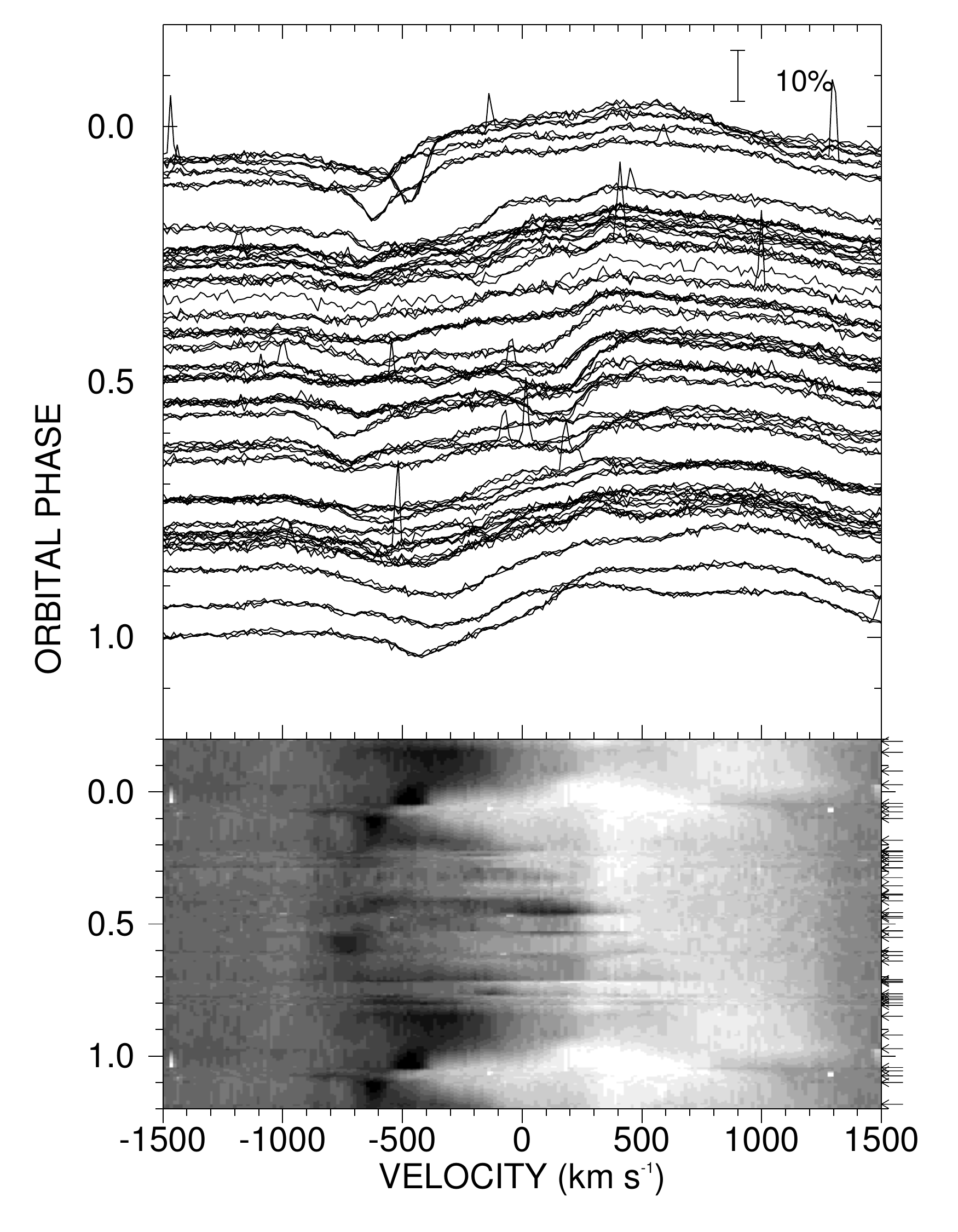}
\endminipage\hfill
\minipage{0.32\textwidth}
  \includegraphics[width=\linewidth]{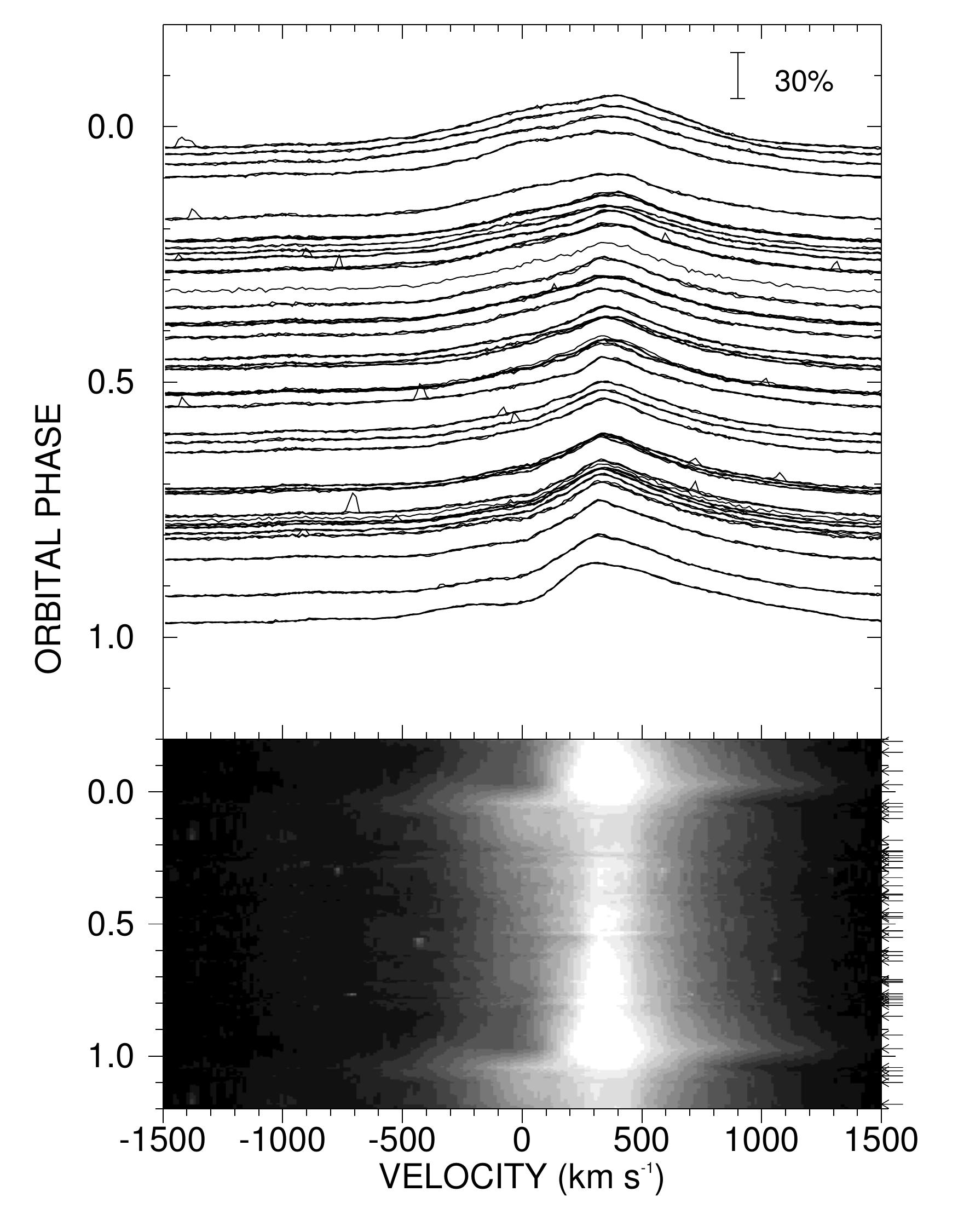}
\endminipage\hfill
\minipage{0.32\textwidth}%
  \includegraphics[width=\linewidth]{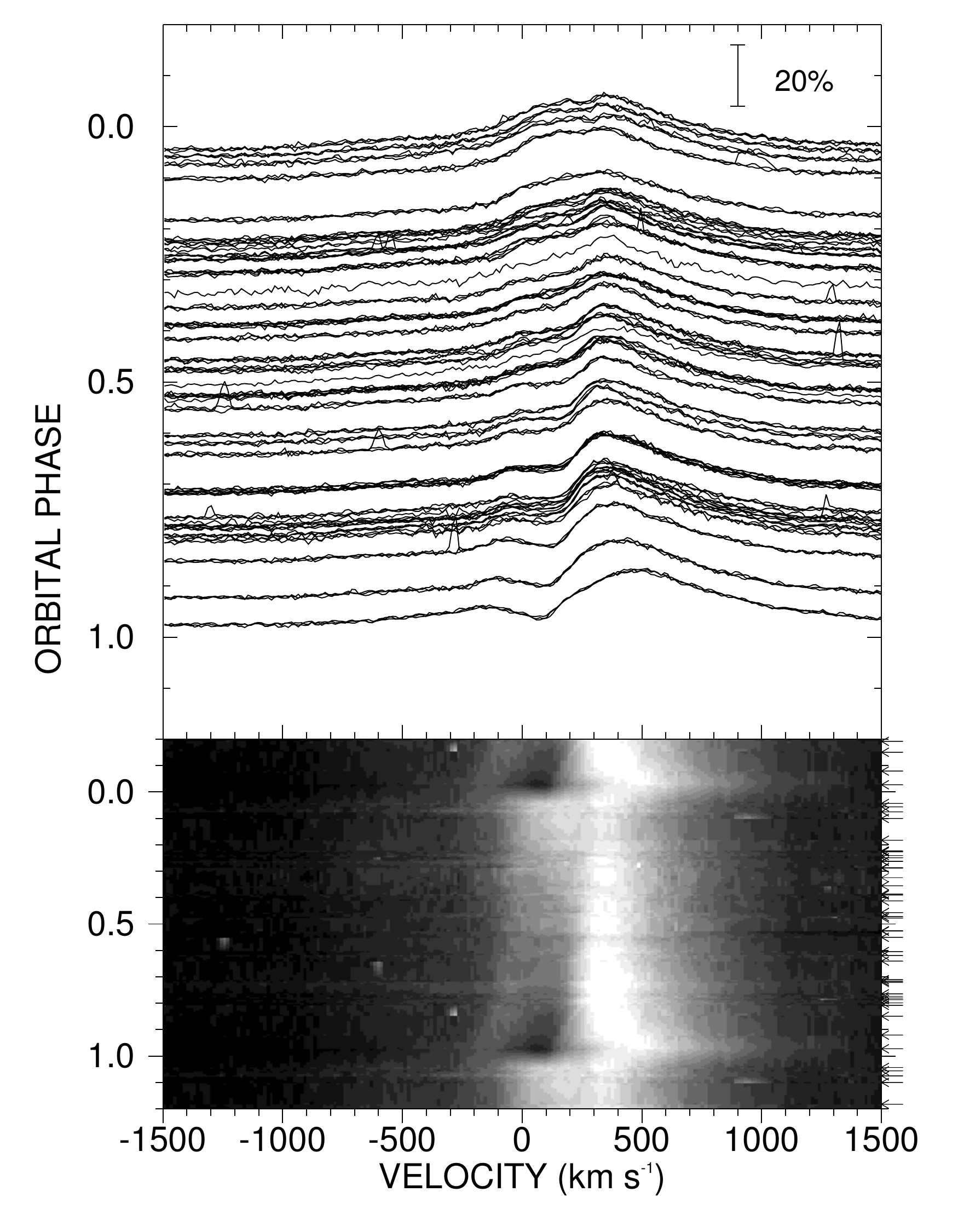}
\endminipage
\caption{Dynamic FLAMES spectra of He\,{\sc i}\,$\lambda 4471$ (left), H$\gamma$ (center), He\,{\sc ii}\,$\lambda 4200$ (right)}
\label{fig:WWC-grayscale}
\end{figure*}

Figure\,\ref{fig:WWC-grayscale} shows dynamic spectra calculated for three
prominent lines: He\,{\sc i}\,$\lambda 4471$, H\,$\gamma$, and He\,{\sc ii}\,$\lambda 4200$. 
The He\,{\sc i} image is especially striking. There is a clear pattern of emission excess  traveling from $\approx -600\,$\kms at $\phi \approx 0$
to $\approx 300$\,\kms at $\phi \approx 0.7$, and back again. This velocity amplitude clearly does not stem from the motion of the stars, 
which trace a different RV pattern and move at amplitudes of $\approx 100\,$\kms. In fact, the emission pattern is fully consistent with a rotating WWC cone, 
as suggested by \citet{Luehrs1997}.  Also interesting are the two strong absorption dips seen close to periastron, which likely occur when the cone arms tangentially 
sweep along the line of sight, thereby instantaneously increasing the optical depth.

To proceed with a more quantitative analysis of the WWC zone, we follow the simple argumentation by \citet{Hill2000}. 
Typically, very strong emission lines are needed for a quantitative analysis 
of WWC. However, our FLAMES spectra do not contain these lines. The most suitable line for the analysis is found to be He\,{\sc i}\,$\lambda 4471$. Due to 
its weakness, only a rough analysis can be made here. The basic model predicts the dependence of the full width half maximum (FW) of the emission excess profile 
and its mean RV as a function of orbital phase $\phi$. $\text{FW}(\phi)$ and $\overline{\text{RV}}(\phi)$ can be written as

\begin{align}
\label{eq:WWC}
\begin{split}
 \text{FW}(\phi) &= C_1 + 2\,v_\text{str}\,\sin \theta \sqrt{1 - \sin^2 i \cos^2 \left(\phi - \delta \phi\right)}
\\ 
 \overline{\text{RV}}(\phi) &= C_2 + v_\text{str} \cos \theta \sin i \cos \left(\phi - \delta \phi\right),
\end{split}
\end{align}
where $C_1$ and $C_2$ are constants, $v_\text{str}$ is the streaming velocity of the shocked gas, $\theta$ is the opening angle of the cone, $i$ is the orbital 
inclination, and $\delta \phi$ is a phase shift introduced due to Coriolis forces \citep[see figure 6 in][]{Hill2000}. 

An unbiased measurement of FW and RV could not be performed because of the low S/N of the line. Instead, the blue and red ``edge'' velocities
of the emission excess, $v_\text{b}$ and $v_\text{r}$,  were 
estimated directly from the gray-scale plot shown in Fig.\,\ref{fig:WWC-grayscale}, from which FW and $\overline{\text{RV}}$ were calculated 
via $\text{FW}(\phi) = v_\text{r}(\phi) - v_\text{b}(\phi)$ and $\overline{\text{RV}}(\phi) = 0.5\left( v_\text{b}(\phi) + v_\text{r}(\phi) \right)$. 
The stream velocity can be deduced from the position of the strong absorption dips around $\phi \approx 0$, seen at approximately $600\,$\kms. Accounting for the systemic 
velocity of the system ($\approx 300\,$\kms), we fix $v_\text{str} = 900$\,\kms. The value found here is slightly lower than the terminal velocity of the primary ($v_\infty \approx 1200\,$\kms), as is expected.

\begin{figure}[!htb]
\centering
  \includegraphics[width=\columnwidth]{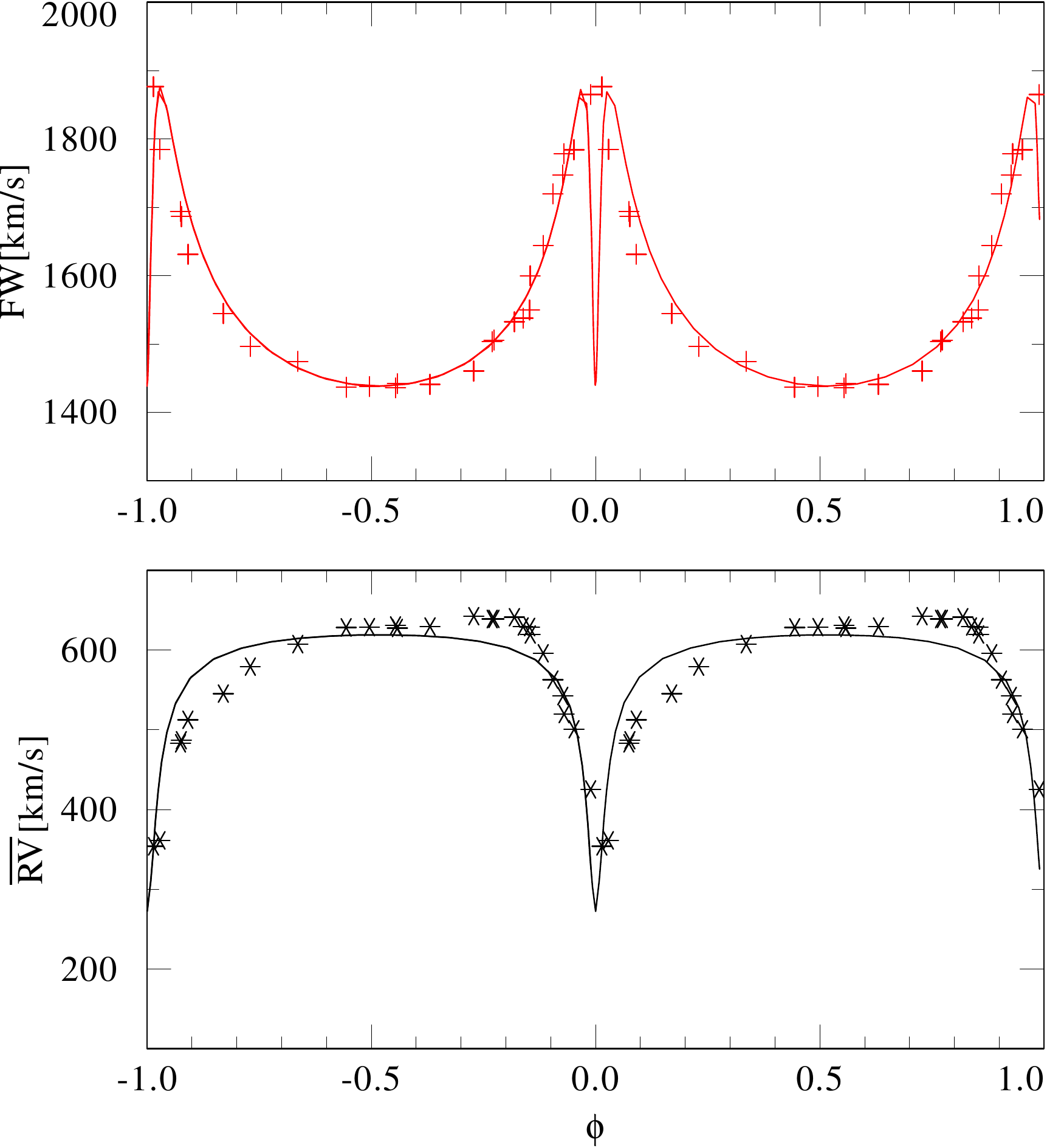}
  \caption{Measured FW($\phi$) and $\overline{\text{RV}}(\phi)$ of the WWC emission excess profiles compared to their best-fitting models.} 
\label{fig:WWCmodel-fit}
\end{figure}

Having fixed $v_\text{str}$, we look for a set of parameters $C_1, C_2, \theta, i,$ and $\delta \phi$ which minimize $\chi^2$. For this purpose, a standard 
python routine (lmfit) is used. Our measurements of FW and $\overline{\text{RV}}$, compared with the best-fitting solutions, are shown in Fig.\,\ref{fig:WWCmodel-fit}. We find 
$C_1 = 150\pm30\,$\kms, $C_2 = 450\pm20$\,\kms, $\theta = 73\pm6^\circ$, $i = 40\pm5^\circ$, and $\delta \phi = 8\pm3^\circ$. The errors given 
here are strongly underestimated, as the true error lies in the measurement technique of the velocities and the simplified model.

It is immediately apparent that both our polarimetric analysis as well as the WWC analysis deliver almost identical inclinations. Admittedly, the value obtained here 
is biased by the method of measuring $v_\text{b}$ and $v_\text{r}$, and should therefore only be considered a further confirmation, rather as an independent 
derivation of $i$. It is also interesting to consider the opening angle $\theta$. \cite{Usov1995} showed that the opening angle can be calculated from

\begin{equation}
 \theta = 2.1 \left(1 - \frac{\eta^{2/5}}{4}\right) \eta^{1/3},
\label{eq:theta}
\end{equation}
where $\eta = \dot{M}_2 v_{\infty,2} / \left( \dot{M}_1 v_{\infty, 1}\right)$ is the wind momentum ratio of both companions. Adopting our
derived values from Table\,\ref{tab:specan}, we find $\eta = 0.48$, which yields $\theta = 76^\circ$, in agreement with our results. 

WWC may also manifest itself via powerful X-ray emission 
\citep[][and references therein]{rauw2015}. However, the presence of strong 
X-ray emission is not a necessary attribute of a colliding wind binary. 
\citet{Oskinova2005} demonstrated that, on average,  the ratio between 
stellar bolometric and X-ray luminosity ($\log L_{\rm X}/L_{\rm bol} \approx -7$) is similar among Galactic massive binary and single stars, not without exceptions 
\citep[e.g.,\ Eta Car,][]{Corcoran1995}. 
R145 was detected by {\em Chandra} X-ray observatory \citep[X-ray source 
designation CXOU J053857.06-690605.6,][]{Townsley2014}. The observations were taken over a period of 9 days
around $T = 53760.7$\,[MJD], corresponding to $\phi \approx 0.75$. 
Using $E_\text{B - V}$ from 
Table\,2 to estimate the interstellar neutral hydrogen column density, the 
observed count rate, and the median energy of the  X-ray photons 
\citep{Townsley2014}, the X-ray luminosity of R145 in the 0.2-12\,keV band 
is $L_{\rm X}\approx 2\times 10^{33}$\,erg\,s$^{-1}$. This corresponds to\footnote{Note that we compare with the total bolometric luminosity of the system because both stars 
are expected to intrinsically emit X-rays.} 
$\log L_{\rm X}/L_{\rm bol,tot}\approx -6.9$. Thus, R145 is not an especially 
luminous X-ray source, albeit it may be somewhat harder than a typical single 
star, as observed in other massive binaries  \citep[e.g.,][]{naze2011}. 
Overall, the X-ray luminosity of R145 is similar to that of other detected  
massive stars in the LMC \citep[e.g.,][]{naze2014} 

The components of R145 follow a highly eccentric orbit. Therefore, modulations of the 
X-ray emission with orbital phases are expected. Previous snap-shot observations 
are not suitable for detecting such orbital modulations. Dedicated 
monitoring X-ray observations of R145 should provide the required information about 
energetic processes in its interacting stellar winds.

\section{The evolution of the system}
\label{sec:disc}

We now exploit the rich information derived here to constrain the evolutionary 
path of R145.
Unfortunately, despite using high-quality data in this study, 
the derived orbital masses suffer from large uncertainties (cf.\ Table\,\ref{tab:orbpar}). This
is mainly due to the small inclination angle $i=39^\circ$, at which even a modest formal error 
of $6^\circ$ translates to an error of $\approx 50\%$ in the mass.  Moreover, due to non-linear biases, 
the value of $i$ obtained here is likely overestimated. Another hindrance is that the FLAMES spectra
poorly cover the periastron passage (see Fig.\,\ref{fig:RVfit}), and so further monitoring would be 
desirable.
Nevertheless, the masses of both components could be derived to an unprecedented precision, 
and set important constraints on the system.


The first question that comes to mind is whether 
the stars in this system have interacted in the past via mass-transfer. 
Evaluating the Roche lobe radii via the 
Eggelton approximation \citep{Eggleton1983} using
the semi major axis $a$, one finds $R_\text{RLOF,1} \approx R_\text{RLOF,2} \approx 360\,R_\odot$.
At closest approach (periastron), 
the distance between the stars is $\left( 1- e\right)\,a$, and the Roche lobe radii would be 
$R_\text{RLOF,1} \approx  R_\text{RLOF,2} \approx 80\,R_\odot$. 
Thus, with radii of $20-30\,R_\odot$  (cf.\ Table\,\ref{tab:specan}),
the stars are safely within their Roche lobes, even at closest approach.

This, however, does not imply that the system had not interacted in the past. Although the primary 
is likely still core H-burning,
it cannot be excluded that the primary exhibited 
larger radii in the past. How compact the primary was throughout its evolution is strongly related to how homogeneous 
it was. Stars undergoing quasi-homogeneous evolution (QHE) tend to maintain much higher temperatures throughout their evolution 
and therefore remain relatively compact \citep[e.g.,][]{Brott2011}. Homogeneity is typically enhanced in stellar evolution 
codes by adopting large initial rotation velocities, which induce chemical mixing \citep{Meynet2005, Heger2000}. Very massive stars may also be 
close to homogeneous simply due to their large convective cores and strong mass-loss rates 
\citep[e.g.,][]{Graefener2011, Vink2015}.
If the primary underwent QHE, mass-transfer was likely avoided in the system. Otherwise, mass-transfer would have occurred in 
the system. The fact that the system is highly eccentric 
is indicative that no mass-transfer has occurred, since RLOF tends to efficiently circularize an orbit \citep{Hurley2002}. 

\subsection{ Comparison with single star tracks}

To gain more insight on the evolutionary course of the system, 
we compare the observables derived here to a set of evolution tracks calculated for single stars. These 
tracks are valid as long as the stars do not interact during their lifetime.
We use tracks calculated by \citet{Brott2011} and \citet{Koehler2015}
for initial masses in the range $5 \leq M_\text{i} \leq 500\,M_\odot$ and 
initial rotational velocities $0 \leq v_\text{rot, i} \lesssim 500\,$\kms 
at a metallicity of $Z = 0.0047$, using the Bonn Evolutionary Code  ("BEC" tracks hereafter), as well as tracks calculated with the BPASS\footnote{bpass.auckland.ac.nz} (Binary 
Population and Spectral Synthesis) code by
\citet{Eldridge2011} and \citet{Eldridge2012} for homogeneous and non-homogeneous single stars with $5 \leq M_\text{i} \leq 150\,M_\odot$ 
and $Z = 0.004$ (``BPASS'' tracks hereafter). 

Finding the initial parameters and age which best reproduce the properties of both components according to the BEC tracks 
is done most easily with the 
BONNSAI\footnote{The BONNSAI web-service is available at www.astro.uni-bonn.de/stars/bonnsai} Bayesian statistics tool \citep{Schneider2014}.
The disadvantage of the BEC tracks is that they, unlike the BPASS tracks, do not include post core H-burning phases.
While the secondary is almost certainly core H-burning given 
its spectral type, this cannot be considered certain for the WR primary, although 
its properties and spectral type 
imply that it is likely  core H-burning  as well \citep[e.g.,][]{Hainich2014}.

\begin{figure*}
\centering
\begin{subfigure}{\columnwidth}
  \centering
  \includegraphics[width=0.95\linewidth]{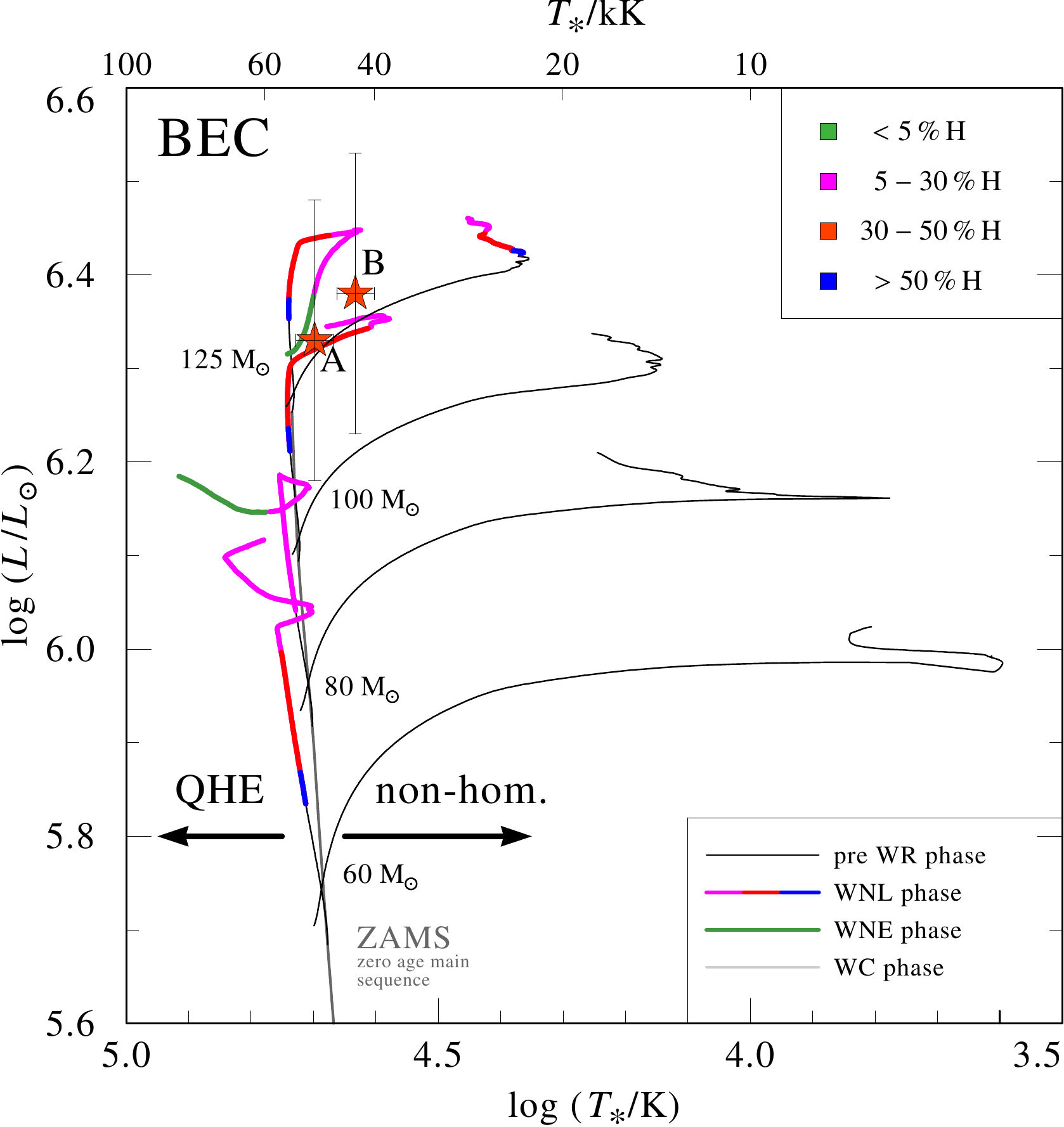}
  \label{fig:sub1}
\end{subfigure}%
\begin{subfigure}{\columnwidth}
  \centering
  \includegraphics[width=0.95\linewidth]{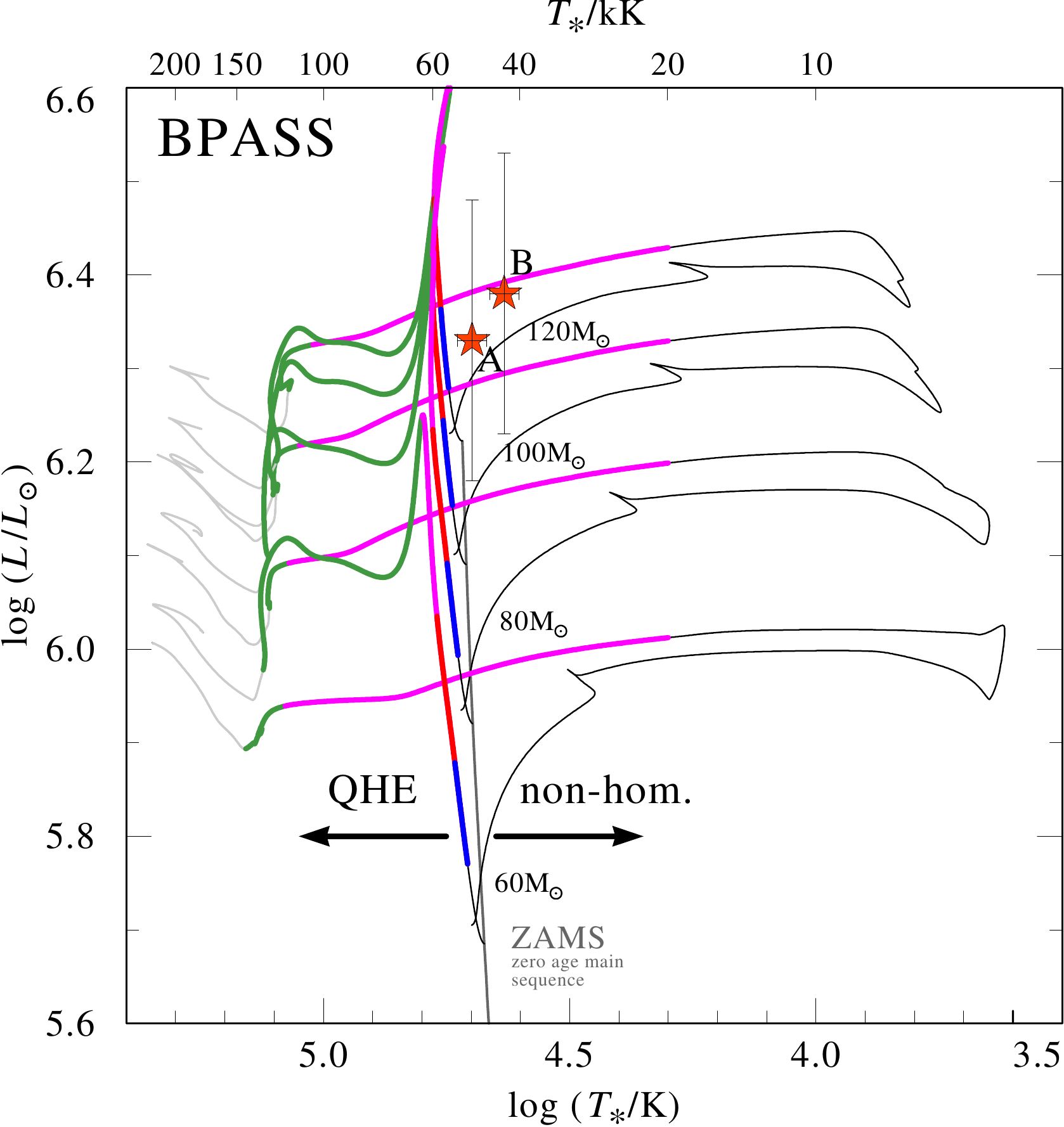}
  \label{fig:sub2}
\end{subfigure}
\caption{The HRD positions of the primary (A) and secondary (B) components of R145 
compared to BEC (left panel) and BPASS (right panel) evolution tracks calculated for (near-)homogeneous and non-homogeneous evolution for LMC metallicities. The WR phase is defined 
for $X_\text{H} < 0.65$ and $T_* > 20\,$kK. See text for details.}
\label{fig:HRDsin}
\end{figure*}

Fig.\,\ref{fig:HRDsin} shows the Hertzsprung-Russell diagram (HRD) positions of 
the primary (A) and secondary (B) components of R145 compared to a selected 
number of BEC (left panel) and BPASS (right panel) evolution tracks. The colors code the amount of surface 
hydrogen content (see legend). We include 
both QHE models as well as non-homogeneous models. For the BEC 
tracks, QHE is reached via high initial rotation rates; the tracks shown 
in Fig.\,\ref{fig:HRDsin} are calcualted for $v_\text{rot, i} \approx 350\,$\kms. The QHE BPASS tracks assume full homogeneity a priori; rotation is 
not considered in the BPASS code.
Note that the QHE BEC tracks are not fully homogeneous.

\subsubsection{ BEC tracks results}

We first use the BONNSAI tool to find the initial parameters which best reproduce
the observables $T_*, L, X_\text{H}$, and $M_\text{orb}$ of the primary, accounting for the errors as found in this study. 
As could be anticipated, only tracks with large 
initial rotations ($v_\text{rot, i} \gtrsim 350\,$\kms) can reproduce its
HRD position (see left panel of Fig.\,\ref{fig:HRDsin}); 
the non-homogeneous tracks terminate at low temperatures and do not return to high 
temperatures because  hydrogen is then exhausted in the core. To obtain 
a consistent set of initial parameters for the secondary, we use the BONNSAI tool again to compare with the secondary's 
observables, but this time, we also use the primary's age (and associated errors), as obtained from the BONNSAI tool.

The resulting initial masses and age (as derived for the primary) are 
shown in Table\,\ref{tab:inpar}. The Table also gives the current masses and hydrogen content of both components as 
predicted from the best-fitting evolutionary track. The initial rotations obtained by the BONNSAI tool for 
the primary and secondary are $v_\text{rot, i} = 410$ and $340\,$\kms, respectively, while the predicted current rotational velocities are 
$240$ and $260\,$\kms, marginally consistent with the upper bounds given in Table\,\ref{tab:specan}. Since the non-homogeneous 
BEC models do not reproduce the HRD positions of the system's components, we give only the corresponding QHE solution in Table\,\ref{tab:inpar}.

\subsubsection{ BPASS tracks results }

A similar procedure is performed with the BPASS tracks. We use a $\chi^2$ minimization algorithm 
to find the best-fitting homogeneous and non-homogeneous tracks and ages which reproduce the properties 
of the primary (see eq.\ 3 in \citealt{Shenar2016}). Once a track and age for the primary is inferred,
we repeat the procedure for the secondary, adopting the primary's age an associated error estimate which is 
based on the grid spacing. The corresponding initial parameters, ages, and current mass and surface hydrogen content, are given 
in Table\,\ref{tab:inpar}. Because the BPASS tracks cover the whole evolution of the star, appropriate solutions can be found for the 
non-homogeneous case as well (see also right panel of Fig.\,\ref{fig:HRDsin}). In Table\,\ref{tab:inpar}, we also give the maximum 
radius $R_\text{max, 1}$ reached by the primary  throughout its evolution. This should give an indication to whether or not the primary 
has exceeded its Roche lobe radius in the past.

\subsubsection{ Indication for QHE }

Both the BEC tracks and the BPASS tracks imply very similar initial parameters and ages in the QHE case for both components.  
The tracks reproduce the observables reasonably well (compared to the errors), but the current masses predicted by the evolutionary 
tracks ($\approx 80-90\,M_\odot$) are larger than the orbital masses 
derived here ($\approx 55\,M_\odot$). Such masses would be obtained at an inclination of $\approx 33^\circ$, which is roughly consistent with our formal 
error on $i$ given in Table\,\ref{tab:orbpar}. The QHE scenario would therefore suggest that the actual masses are $\approx 80-90\,M_\odot$ per component. 

For the non-QHE scenario, only the BPASS tracks can place meaningful constraints. In this scenario, the properties of the 
primary are reproduced when the evolution tracks "return" from the red to the blue, and He-core burning initiates. This scenario 
is consistent with much lower current masses, closer to those derived here.
However, significant discrepancy is obtained for the hydrogen content. More importantly, a comparison between the maximum radius 
reached by the primary and the Roche lobe implies that, assuming the non-homogeneous tracks, 
the primary overfilled its Roche lobe in the past. Note that the separation increases with time due to mass-loss, 
making mass-transfer inevitable in the non-homogeneous case.  In this scenario, binary interaction 
therefore has to be accounted for.

\renewcommand{\arraystretch}{1.2}
\setlength{\tabcolsep}{2mm}

\begin{table}[!htb]
\scriptsize
\small
\tabcolsep=0.1cm
\caption{Comparison with best-fitting evolution tracks}
\label{tab:inpar}
\begin{center}
\begin{tabular}{l | c | c | c | c}
\hline      
& BEC  & \multicolumn{3}{c}{BPASS}  \\ 
& QHE\tablefootmark{a}       & non-hom.\ & QHE\tablefootmark{b} & Binary\tablefootmark{c} \\ 
\hline
$M_\text{i,1}$ [$M_\odot$]      & $105\pm20$          & $100\pm20$     & $100\pm20$  & $120\pm20$       \\
$M_\text{i,2}$ [$M_\odot$]      & $91\pm15$           & $90\pm20$      & $90\pm20$   & $80\pm20$             \\
$M_\text{cur,1}$ [$M_\odot$]    & $77^{+30}_{-15}$    &  $48\pm10$     & $96$        & $58$               \\
$M_\text{cur,2}$ [$M_\odot$]    & $78^{+20}_{-10}$    &  $52\pm10$     & $85$        & $113$                \\
Age [Myr]                       & $2.3\pm0.4$         &  $3.1\pm0.3$   & $2.1\pm0.4$ & $2.8\pm0.4$          \\
$X_\text{H,1}$ (mass fr.)       & $0.3\pm0.1$         &  $0.1\pm0.1$   & $0.4\pm0.1$ & $0.1\pm0.1$          \\
$X_\text{H,2}$ (mass fr.)       & $0.74^{+0}_{-0.25}$ &  $0.1\pm0.1$   & $0.5\pm0.1$ & -                 \\
$R_\text{max, 1}$ [$R_\odot$]   & $ 40$               & $1500$         & $20$        & -                \\
\hline
\end{tabular}
\tablefoot{
Initial parameters and predictions of best-fitting evolution tracks calculated for single stars experiencing homogeneous/non-homogeneous evolution.
\tablefoottext{a}{QHE in the BEC tracks is reached via an initial rotation velocity of $\approx 350\,$\kms for both components, which best-reproduces the observables.}
\tablefoottext{b}{These tracks are fully homogeneous a-priori; rotation is not accounted for in the BPASS tracks.}
\tablefoottext{c}{The tracks are identical to the non-homogeneous tracks, but include mass-transfer. The initial period defining the best-fitting track is $P_\text{i} = 100\,$d.}
}
\end{center}
\end{table}

\subsection{ Binary tracks }

We now use set of tracks calculated with version 2.0 of the BPASS code \citep{Eldridge2008} for $Z = 0.004$ which are non-homogeneous and account 
for mass-transfer. Each track is defined by an
initial period $P_\text{i}$, an initial mass ratio $q_\text{i} = M_\text{i, 2} / M_\text{i, 1}$, and an initial mass for the primary $M_\text{i,1}$, 
calculated at intervals of 0.2 on $0 < \log P\,[\text{d}] < 4$, 0.2 on $0 < q_\text{i} < 0.9$, and $10-20\,M_\odot$ on $10 < M_\text{i,1} < 150\,M_\odot$. 
The tracks do not include the hydrogen abundance of the secondary, so only a comparison with the primary's $X_\text{H}$ is possible.

To find the track which best reproduces the observables, we use a $\chi^2$ minimization algorithm (see eq.\ 4 in \citealt{Shenar2016}). 
We account here for $T_*, L, M_\text{orb}$ for both components, $X_\text{H}$ for the primary, the current period $P$, and the current mass-ratio 
$q = M_2 / M_1 = K_1 / K_2$. Note that we include the mass ratio because it has a much smaller formal error ($q = 1.01 \pm 0.07$), as opposed to the 
actual masses. 

The parameters of the best fitting binary track are given in the last column of Table\,\ref{tab:inpar}.  
Even the best fitting track results in a mass ratio of almost $2$. The reason is that in all relevant tracks, RLOF 
from the primary to the secondary occurs, which tends to result in mass ratios significantly different than $1$. 
The binary track also fails to reproduce the large hydrogen mass fraction inferred for the primary. 
Binary evolution tracks therefore do poorly in reproducing the system's observables.

\subsection{ Outlook}

To summarize, it appears that the system has evolved quasi-homogeneously, similarly to HD 5980 \citep{Koenigsberger2014}.
This would suggest current 
masses of $\approx 80\,M_\odot$ and initial masses of $M_1 \approx 105\,M_\odot$ and $M_2 \approx 90\,M_\odot$.  The current 
generation of evolution models can attain QHE only via rapid initial rotation $v_\text{i} \gtrsim 350\,$\kms.
Admittedly,
one may argue that it is unlikely for both stars to be born with such high initial rotations \citep[e.g.,][]{Ramirez2015}.
A possible resolution could lie in tidal interaction during periastron passage. The high eccentricity of the system 
yields a small separation between the components during periastron, which in turn may imply significant tidal mixing during periastron passage.
As noted above, homogeneity can also be obtained by virtue of the large convective cores 
and strong mass-loss \citep{Graefener2011} of massive stars. Hence, the rotation which is needed for the BONN tracks may serve as a proxy for QHE rather 
than point at the actual physical mechanism responsible for homogeneity.


Assuming QHE indeed took place, there is no obvious reason to expect that the components would interact via RLOF in the future. 
This would be the case if the secondary would overfill its Roche lobe, which is currently hard to predict.  If the components will indeed avoid 
interaction in the future, the system will likely evolve into a wind-fed high mass X-ray binary.
With the help of fortunate kicks during core-collapse, the system 
could become close enough to merge within a Hubble time, emitting a gravitational wave event like those recently 
observed with LIGO \citep{Abbott2016, Marchant2016}.

\section{Summary}
\label{sec:summary}

We have performed an exhaustive analysis of the very massive system BAT99 119 (R145) in the LMC.
Using high-quality FLAMES spectra, 
we detected and resolved for the first time lines from the secondary component and derived
a first SB2 orbital solution for the system. 
The composite FLAMES 
spectral were disentangled to the constituent spectra of both components, and a spectral analysis was performed to derive 
the physical parameters of the components. 
This enabled us to confirm the primary's spectral type as WN6h, and to infer for the first 
time a spectral type for the secondary: O3.5~If*/WN7.
A polarimetric analysis, as well as a WWC analysis, helped constrain the 
inclination of the system. Finally, a comparison with evolution tracks was conducted.

The system was previously speculated to host the most 
massive stars known ($M_1 > 300\,M_\odot$, S2009). 
From our orbital + polarimetric analysis, we derive 
$q = M_2 / M_1 = 1.01\pm0.07$ and masses $M_1 \approx M_2 \approx 55^{+40}_{-20}\,M_\odot$. Thus, although the masses 
suffer from large uncertainties, we can exclude masses larger than $100\,M_\odot$ in the system. 


We find clear evidence for WWC in the system. Interestingly, the signature of WWC is only clear in low-ionization transitions.
We could only perform a rough quantitative spectroscopic analysis of the WWC spectral features because of the absence 
of very strong lines which are affected by WWC. The resulting inclination ($i = 40^\circ$) is consistent with that obtained from polarimetry
 $i = 39^\circ$), and the half opening 
angle ($\theta = 76^\circ$) is consistent with the mass-loss rates and terminal velocities derived from the spectral analysis. 

A comparison with quasi homogeneous and non-homogeneous 
BEC and BPASS evolution tracks, the latter accounting for mass transfer as well, implies that quasi-homogeneous evolution (QHE) best describes 
the system. In this scenario, the components remain compact throughout their evolution and do not fill their Roche lobes. 
Non-homogeneous evolution would imply mass transfer, and this in turn leads to mass ratios which are very different than 
found here ($\approx 1$),  which is why we can exclude non-homogeneous evolution to a high degree of certainty. 
The high eccentricity found in this study ($e \approx 0.8$) is in line with the fact that the components did not interact via RLOF, which would tend to 
circularize the system.
However, QHE 
is only consistent if the current masses are $\approx 80-90\,M_\odot$, which is roughly the upper limit of our derived orbital masses. In any case, 
the initial masses of the stars are found to be $M_\text{1, i} \approx 105$ and $M_\text{2, i} \approx 90\,M_\odot$.

Future spectroscopic and polarimetric observations are strongly encouraged to obtain more spectral phase coverage during periastron passage, 
which would constrain the orbital fit further and reduce uncertainties. 
A phase coverage of the red optical spectrum, as well as X-ray light curves, would be highly helpful in analyzing the WWC region to a much larger degree 
of accuracy, enabling an accurate derivation of the inclination, and a detailed study of WWC in this important system.

\begin{acknowledgements}
We are grateful for the constructive comments of our referee. 
TS acknowledges the financial support from the Leibniz Graduate School for Quantitative 
Spectroscopy in Astrophysics, a joint project of the Leibniz Institute for Astrophysics Potsdam (AIP) 
and the institute of Physics and Astronomy of the University of Potsdam. 
AFJM is grateful for financial aid from NSERC (Canada) and FQRNT (Quebec).
NDR is
grateful for postdoctoral support by the University of Toledo and by the Helen Luedke Brooks Endowed Professorship.
LAA acknowledges support from the Fundac\~ao de Amparo \`a Pesquisa do Estado de S\~{a}o Paulo - FAPESP (2013/18245-0 and 2012/09716-6). 
RHB thanks support from FONDECYT project No.\ 1140076.
\end{acknowledgements}
\bibliography{literature}

\Online

\begin{appendix}
\section{RV measurements}


\begin{table}[htb!]
\tabcolsep=0.1cm
\small
\caption{RVs for primary (N\,{\sc iv}) and secondary (Si\,{\sc iv}) components}
\begin{center}
\begin{tabular}{c | c | c | c | c}
\hline
 Spectrum & MJD & $\phi$ & N\,{\sc iv} RV [\kms] & Si\,{\sc iv} RV [\kms] \\ 
\hline
1 & 54794.18 & 0.27 & 230 & 302\\ 
2 & 54794.20 & 0.27 & 256 & 286\\ 
3 & 54794.24 & 0.27 & 220 & 298\\ 
4 & 54794.32 & 0.27 & 234 & 300\\ 
5 & 54798.29 & 0.29 & 234 & 288\\ 
6 & 54798.31 & 0.29 & 234 & 298\\ 
7 & 54804.07 & 0.33 & 266 & 288\\ 
8 & 54836.13 & 0.53 & 280 & 284\\ 
9 & 54836.16 & 0.53 & 280 & 278\\ 
10 & 54836.18 & 0.53 & 280 & 274\\ 
11 & 54836.20 & 0.53 & 280 & 274\\ 
12 & 54867.05 & 0.72 & 292 & 264\\ 
13 & 54867.07 & 0.72 & 296 & 260\\ 
14 & 55108.27 & 0.24 & 234 & 296\\ 
15 & 55108.29 & 0.24 & 234 & 298\\ 
16 & 56210.35 & 0.18 & 226 & 308\\ 
17 & 56210.37 & 0.18 & 220 & 306\\ 
18 & 56210.38 & 0.18 & 218 & 302\\ 
19 & 56217.33 & 0.23 & 234 & 300\\ 
20 & 56217.34 & 0.23 & 234 & 300\\ 
21 & 56217.35 & 0.23 & 234 & 294\\ 
22 & 56243.34 & 0.39 & 256 & 288\\ 
23 & 56243.35 & 0.39 & 252 & 300\\ 
24 & 56243.36 & 0.39 & 256 & 292\\ 
25 & 56256.26 & 0.47 & 270 & 278\\ 
26 & 56256.27 & 0.47 & 280 & 272\\ 
27 & 56256.28 & 0.47 & 276 & 268\\ 
28 & 56257.13 & 0.48 & 270 & 268\\ 
29 & 56257.14 & 0.48 & 270 & 278\\ 
30 & 56257.15 & 0.48 & 272 & 280\\ 
31 & 56277.31 & 0.61 & 282 & 268\\ 
32 & 56277.32 & 0.61 & 284 & 266\\ 
33 & 56277.33 & 0.61 & 276 & 262\\ 
34 & 56283.05 & 0.64 & 296 & 258\\ 
35 & 56283.06 & 0.64 & 268 & 298\\ 
36 & 56283.07 & 0.64 & 290 & 258\\ 
37 & 56294.20 & 0.71 & 302 & 258\\ 
38 & 56294.21 & 0.71 & 302 & 264\\ 
39 & 56294.23 & 0.71 & 294 & 252\\ 
40 & 56295.18 & 0.72 & 296 & 254\\ 
41 & 56295.19 & 0.72 & 294 & 258\\ 
42 & 56295.21 & 0.72 & 294 & 260\\ 
43 & 56304.24 & 0.77 & 294 & 250\\ 
44 & 56305.23 & 0.78 & 294 & 252\\ 
45 & 56305.24 & 0.78 & 294 & 244\\ 
46 & 56305.26 & 0.78 & 294 & 260\\ 
47 & 56306.22 & 0.79 & 294 & 238\\ 
48 & 56306.23 & 0.79 & 268 & 276\\ 
49 & 56306.24 & 0.79 & 294 & 246\\ 
50 & 56308.15 & 0.80 & 296 & 242\\ 
\hline
\end{tabular}
\end{center}
\label{tab:RVs}
\end{table}

\addtocounter{table}{-1}

\begin{table}[!htb]
\tabcolsep=0.1cm
\small
\caption{Continued}
\begin{center}
\begin{tabular}{c | c | c | c | c}
\hline
 Spectrum & MJD & $\phi$ & N\,{\sc iv} RV [\kms] & Si\,{\sc iv} RV [\kms] \\ 
\hline
51 & 56308.17 & 0.80 & 296 & 242\\ 
52 & 56308.18 & 0.80 & 280 & 244\\ 
53 & 56316.21 & 0.85 & 296 & 234\\ 
54 & 56316.22 & 0.85 & 296 & 226\\ 
55 & 56316.23 & 0.85 & 310 & 236\\ 
56 & 56347.01 & 0.04 & 222 & 326\\ 
57 & 56347.03 & 0.04 & 232 & 326\\ 
58 & 56347.04 & 0.04 & 222 & 322\\ 
59 & 56349.02 & 0.06 & 206 & 328\\ 
60 & 56349.03 & 0.06 & 208 & 328\\ 
61 & 56349.05 & 0.06 & 204 & 328\\ 
62 & 56352.02 & 0.08 & 224 & 342\\ 
63 & 56352.04 & 0.08 & 268 & 284\\ 
64 & 56352.05 & 0.08 & 214 & 334\\ 
65 & 56356.00 & 0.10 & 268 & 284\\ 
66 & 56356.02 & 0.10 & 192 & 332\\ 
67 & 56356.03 & 0.10 & 202 & 322\\ 
68 & 56571.34 & 0.46 & 264 & 284\\ 
69 & 56571.35 & 0.46 & 272 & 288\\ 
70 & 56571.37 & 0.46 & 406 & 142\\ 
71 & 56571.38 & 0.46 & 270 & 286\\ 
72 & 56582.34 & 0.53 & 280 & 270\\ 
73 & 56582.35 & 0.53 & 280 & 274\\ 
74 & 56582.37 & 0.53 & 280 & 278\\ 
75 & 56586.25 & 0.55 & 280 & 280\\ 
76 & 56586.26 & 0.55 & 272 & 270\\ 
77 & 56586.27 & 0.55 & 280 & 278\\ 
78 & 56597.23 & 0.62 & 236 & 298\\ 
79 & 56597.24 & 0.62 & 276 & 262\\ 
80 & 56597.25 & 0.62 & 280 & 262\\ 
81 & 56620.26 & 0.76 & 294 & 250\\ 
82 & 56620.27 & 0.76 & 294 & 248\\ 
83 & 56620.28 & 0.76 & 294 & 262\\ 
84 & 56627.16 & 0.81 & 294 & 234\\ 
85 & 56627.18 & 0.81 & 294 & 246\\ 
86 & 56627.19 & 0.81 & 338 & 196\\ 
87 & 56645.04 & 0.92 & 256 & 298\\ 
88 & 56645.05 & 0.92 & 340 & 194\\ 
89 & 56645.07 & 0.92 & 338 & 196\\ 
90 & 56653.28 & 0.97 & 414 & 152\\ 
91 & 56653.29 & 0.97 & 408 & 146\\ 
92 & 56653.30 & 0.97 & 406 & 142\\ 
93 & 56693.11 & 0.22 & 234 & 304\\ 
94 & 56693.12 & 0.22 & 238 & 306\\ 
95 & 56693.13 & 0.22 & 240 & 304\\ 
96 & 56697.16 & 0.25 & 236 & 296\\ 
97 & 56697.17 & 0.25 & 236 & 288\\ 
98 & 56697.19 & 0.25 & 236 & 298\\ 
99 & 56703.13 & 0.29 & 252 & 300\\ 
100 & 56703.14 & 0.29 & 252 & 302\\ 
101 & 56703.16 & 0.29 & 246 & 292\\ 
102 & 56714.02 & 0.36 & 252 & 288\\ 
103 & 56714.03 & 0.36 & 252 & 288\\ 
104 & 56714.05 & 0.36 & 252 & 290\\ 
105 & 56719.02 & 0.39 & 256 & 286\\ 
106 & 56719.03 & 0.39 & 256 & 298\\ 
107 & 56719.04 & 0.39 & 268 & 298\\ 
108 & 56723.17 & 0.41 & 268 & 284\\ 
109 & 56723.18 & 0.41 & 268 & 276\\ 
110 & 56723.20 & 0.41 & 268 & 284\\ 
\hline
\end{tabular}
\end{center}
\label{tab:RVs}
\end{table}

\end{appendix} 

\end{document}